\documentclass{LMCS}

\usepackage{stmaryrd}
\usepackage{amssymb}
\usepackage{amsmath}
\usepackage{latexsym}
\usepackage{derivtree}
\usepackage{calrsfs}
\usepackage{url}
\usepackage[matrix,graph,curve,tips]{xy}\SelectTips{cm}{10}

\usepackage[pdftitle={From Proof Nets to the Free *-Autonomous Category},%
pdfauthor={Francois Lamarche and Lutz Strassburger}]{hyperref}

\def\epsilont{\tilde{\epsilon}}
\def\fh{\hat{f}}
\def\gh{\hat{g}}

\def\cA{{\mathcal A}}

\def\CC{{\mathcal C}}

\def\cF{{\mathcal F}}

\def\cL{{\mathcal L}}

\def\Gflat{{G^\flat}}

\def\PN{{\mathbf{PN}}}

\def\id{\mathit{1}}
\def\idh{\hat{\id}}
\def\Hom{{\mathrm{Hom}}}
\def\Obj{{\mathrm{Obj}}}
\def\isom{\cong}

\def\complI{{\complement I}}
\def\complJ{{\complement J}}

\def\loneC{{\lone_\cC}}
\def\lbotC{{\bot_\cC}}

\def\Gnull{{G^\circ}}

\def\Gbn#1{G(#1\lfmark)\lneg}

\def\lone{\mathsf{1}}
\def\lbot{\bot}
\def\ltens{\varotimes}
\def\lcut{\varobar}
\def\lpar{\mathop\bindnasrepma}
\def\lneg{^\bot}
\def\lnegneg{^{\bot\bot}}
\def\limp{\multimap}

\newbox\lbigparbox
\setbox\lbigparbox=\hbox{\Large$\lpar$}
\newbox\lbigtensbox
\setbox\lbigtensbox=\hbox{\Large$\ltens$}

\def\lbigparsymb{\vcenter{\copy\lbigparbox}}
\def\lbigtenssymb{\vcenter{\copy\lbigtensbox}}

\def\lbigpar#1#2{\lbigparsymb_{#1}#2}
\def\lbigtens#1#2{\lbigtenssymb_{#1}#2}

\def\lbigparind#1#2#3{\def\i{_#2}\lbigparsymb_{#1}\bigset{#3\mid#2\in #1}}
\def\lbigtensind#1#2#3{\def\i{_#2}\lbigtenssymb_{#1}\bigset{#3\mid#2\in #1}}

\def\fneg{(-)\lneg}
\def\cutr{\mathsf{cut}}
\def\idr{\mathsf{id}}
\def\exr{\mathsf{ex}}
\def\rr{\mathsf{r}}

\def\grammareq {\mathrel{\raise.4pt\hbox{::}{=}}}%

\def\set#1{\{#1\}}
\def\bigset#1{\big\{#1\big\}}
\def\dset#1{\left\{#1\right\}}

\def\cons#1{\{#1\}}
\def\conhole      {\cons{\enspace}}%

\def\prfnetsymbol{\mathrel{\vartriangleright}}
\def\prfgrfT#1#2{#1\prfnetsymbol#2}
\def\prfgrfD#1#2{%
\begin{array}{c}
#1\\
\triangledown\\
#2
\end{array}
}
\def\prfgrf#1#2{\mathchoice
  {\prfgrfD{#1}{#2}}%
  {\prfgrfT{#1}{#2}}%
  {\prfgrfT{#1}{#2}}%
  {\prfgrfT{#1}{#2}}%
}

\def\prfneteq#1#2{[#1\prfnetsymbol#2]_\prfequ}
\def\prfnet#1#2{[#1\prfnetsymbol#2]}

\def\unitfree#1#2{\pi_{\prfgrfT{#1}{#2}}}
\def\prfequ{\sim}
\def\prfred{\rightarrow}
\def\redprf{\leftarrow}

\def\fcomp{\circ}

\def\lfmark{^\star}
\def\lneglfmark{^{\lbot\star}}
\def\lfmarklneg{^{\star\lbot}}

\newcommand{\ie}{i.e.,\ }
\newcommand{\eg}{e.g.,\ }

\newcommand{\Wolg}{Without loss of generality}

\newcommand{\Bwoc}{By way of contradiction}

\def\quadfs {\rlap{\rm\quad.}}%
\def\quadcm {\rlap{\rm\quad,}}%
\def\quand {\quad\mbox{and}\quad}%
\def\qquand {\qquad\mbox{and}\qquad}%

\newcommand{\mydisplaybox}[2][-10pt]{
  \newdimen\mydisplaywidth
  \mydisplaywidth=\displaywidth
  \advance\mydisplaywidth by#1  
  \hbox to\mydisplaywidth{\hfil$#2$\hfil}
  }

\def\twodisplay#1#2{%
  \newdimen\mydisplaywidth
  \mydisplaywidth=\displaywidth
  \advance\mydisplaywidth by-10pt 
  \begin{array}{c}
    \hbox to\mydisplaywidth{$#1$\hfil}\\
    \hbox to\mydisplaywidth{\hfil$#2$}
  \end{array}
}

\let\dfn=\emph

\theoremstyle{plain}

\newtheorem{theorem}{Theorem}[subsection]

\newtheorem{lemma}[theorem]{Lemma}
\newtheorem{proposition}[theorem]{Proposition}

\theoremstyle{definition}

\newtheorem{remark}[theorem]{Remark}
\newtheorem{definition}[theorem]{Definition}
\newtheorem{observation}[theorem]{Observation}
\newtheorem{para}[theorem]{}

\def\sone{\strut\lone}
\def\sbot{\strut\lbot}
\def\pbot{\phantom{\strut\lbot}}
\def\sa{\strut a}
\def\sna{\strut a\lneg}
\def\ptens{\phantom{\strut\ltens}}

\def\simpleheight{2.2}
\def\doubleheight{3.2}
\def\tripleheight{4.5}
\def\quadrupleheight{5.5}
\def\quintupleheight{6.5}
\def\sextupleheight{7.5}
\def\lowerheight{-1}
\def\lowerhalf{-0.5}

\newenvironment{introdia}{
\begin{xy} 
0;<1em,0ex>:<0em,4ex>::
(0,1)="a"*{\sone}, (2,1)="b"*{\sbot}, (4,1)="c"*{\sbot},
(6,1)="d"*{\sone}, (8,1)="e"*{\sbot}, (10,1)="f"*{\sbot}, (12,1)="g"*{\sone},
(3,0)="h"*{\ltens},(9,0)="i"*{\ltens},
"b"*{\pbot};"h"*{\pbot} **\dir{-},
"c"*{\pbot};"h"*{\pbot} **\dir{-},
"e"*{\pbot};"i"*{\pbot} **\dir{-},
"f"*{\pbot};"i"*{\pbot} **\dir{-},
}
{\end{xy}}

\def\matrcol{1.8ex}
\def\matrrow{0ex}
\def\matrrowa{0.5ex}

\def\mclap#1{\hbox to 0.5ex{\hss#1\hss}}

\def\mnix{\mclap{}\phantom{\clap{$\oplus$}}\mat{\;}}
\def\mnoat{\mat{\;}}
\def\mtens#1#2{\mclap{$\smash{\ltens}$}\phantom{\clap{$\oplus$}}
  \ar@{-}[#1]\ar@{-}[#2]}
\def\mcut#1#2{\mclap{$\smash{\lcut}$}\phantom{\clap{$\oplus$}}
  \ar@{-}[#1]\ar@{-}[#2]}
\def\mpar#1#2{\mclap{$\smash{\lpar}$}\phantom{\clap{$\oplus$}}
  \ar@{-}[#1]\ar@{-}[#2]}
\def\marc#1{\phantom{\clap{$\oplus$}}\ar@{-}[#1]}
\def\mup{\ar@{-}[u]}
\def\mdown{\ar@{-}[d]}
\def\mbot{\strut\lbot}
\def\mone{\strut\lone}
\def\mat#1{\strut#1}
\def\matn#1{\strut\rlap{\smash{\hbox{$#1\lneg$}}}\phantom{#1}}

\def\diarow{5ex}

\newcommand\formalsystem{\mathcal{S}}
\newcommand\categoricalaxioms{\mathcal{T}}
\newcommand\myst{\tau}

\newcommand\wht{\circ}
\newcommand\blk{\bullet}
\newcommand\seq{\vdash}

\def\doi{2 (4:3) 2006}
\lmcsheading%
{\doi}
{1--44}
{}
{}
{Nov.~\phantom{0}3, 2005}
{Oct.~\phantom{0}5, 2006}
{}

\begin{document}

\title{From Proof Nets to the Free $*$-Autonomous Category}

\author[F.~Lamarche]{Fran\c{c}ois Lamarche\rsuper{a}}
\address{{\lsuper{a}}LORIA \& INRIA-Lorraine\\
Projet Calligramme\\
615, rue du Jardin Botanique\\
54602 Villers-l{\`e}s-Nancy\\
France}
\email{lamarche@loria.fr}
%
\author[L.~Stra\ss burger]{Lutz Stra\ss burger\rsuper{b}}
\address{{\lsuper{b}}INRIA-Futurs\\
Projet Parsifal\\
\'Ecole Polytechnique, Laboratoire d'Informatique (LIX)\\
Rue de Saclay\\
91128 Palaiseau Cedex\\
France}
\email{lutz@lix.polytechnique.fr}
\thanks{{\lsuper{b}}This research has been carried out while the second author had
an INRIA post-doc position at the LORIA in Nancy, France.}

\keywords{multiplicative linear logic, proof nets, $*$-autonomous categories,
coherence problems}
\subjclass{F.4.1}

\begin{abstract}
  In the first part of this paper we present a theory of proof nets
  for full multiplicative linear logic, including the two units. It
  naturally extends the well-known theory of unit-free multiplicative
  proof nets. A linking is no longer a set of axiom links but a tree
  in which the axiom links are subtrees. These trees will be
  identified according to an equivalence relation based on a simple
  form of graph rewriting. We show the standard results of
  sequentialization and strong normalization of cut elimination.  In
  the second part of the paper we show that the identifications
  enforced on proofs are such that the class of two-conclusion proof
  nets defines the free $*$-autonomous category.
\end{abstract}

\maketitle


\section*{Introduction}

The interplay between logic and category theory is fascinating because
it is rich, bidirectional and non-trivial. There is more to this
non-triviality than the fact that
\begin{quote}
  a proof of a statement like ``the logical system \(\formalsystem\)
  corresponds to the set of categorical axioms
  \(\categoricalaxioms\)'' is always a non-trivial task.
\end{quote}\smallskip

\noindent In addition there will very often be discrepancies between
the abstract categorical axiomatization and the actual properties of
the syntactical objects that are used by proof theorists. And if a
denotational semantics is found for \(\formalsystem\), it is more
likely to follow the categorical directives than the syntactical ones.
These discrepancies are the source of creative tensions.

For instance many logical constructions can be expressed in terms of
adjunctions, and ordinary adjunctions give rise to two ``triangular''
equations, which can be called (very roughly) unit and co-unit. But
syntactical considerations often give a real significance to one of
them but not to the other.  A standard example is the lambda calculus,
where the co-unit equation is \(\beta\)-reduction, and the unit one is
the \(\eta\)-rule.  Nobody would suggest that the latter is more
important than the former, and proof theorists would most often rather
not deal with the \(\eta\)-rule, because it makes normalization much
harder, if not impossible. But it is not easy at all to construct a
denotational semantics that does not obey the \(\eta\)-rule, although
it can be done~\cite{hayashi:85}.

As another example of tension, if a poset can be used to embody
provability---\(A \le B \) means I can prove B if I assume
\(A\)---replacing that poset by a category will allow us to
\emph{name} proofs, and to single one out by a map \(f\colon A \to
B\). But then composition of proofs (for syntacticians:
cut-elimination) will have to be associative. This happens rather
naturally with natural deduction systems, less so with the sequent
calculus, where some quotienting has to be done. Thus category theory
furnishes critical tools to test proof theory, a set of external
ideals by which it can be judged.  But if some categorical criterion
is not obeyed by the syntax, this does not mean that syntax is
automatically wrong. Perhaps it is the categorical formulation that
needs to be refined. Tensions can be resolved in more than one way.

Naturally this idea of naming proofs ``correctly'' has been around
long before categories were invented. For a long time logicians have
been aware of the need to determine, given a formal system
\(\mathcal{S}\) and two proofs of a formula \(A\) in that system, when
these two proofs are, or name ``the same'' proof. As a matter of fact
this was already a concern of Hilbert when he was preparing his famous
lecture of 1900~\cite{thiele:03}.  This problem has taken more
importance during the last few years, because many logical systems
permit a close correspondence between proofs and programs.

In a formalism like the sequent calculus (and to a lesser degree,
natural deduction), it is oftentimes very easy to see that two
derivations $\pi_1$ and $\pi_2$ should be identified because $\pi_1$
can be transformed in to $\pi_2$ by a sequence of rule permutations
that are obviously trivial.  It is less immediately clear \emph{in
  general} what transformations can be effected on a proof without
changing its essence. Here the categorical ideals are very helpful,
providing criteria for the identification of proofs that are simple,
general and unambiguous. But they are sometimes too strong, as
happens~\cite{lambek:scott:86,girard:LC} for classical
logic\footnote{perhaps it is better to say: the currently held
  conception of proofs in classical logic}\ldots\ another case of
creative tension, which puts evolutionary pressure on both category
theory and proof theory.

The advent of linear logic marked a significant advance in that quest
for a good onomastics of proofs.  In particular the multiplicative
fragment of linear logic (MLL) comes equipped with an extremely
successful theory of proof identification: not only do we know exactly
when two sequent proofs should be identified (the allowed rule
permutations are described in~\cite{lafont:95}), but there is a class
of simple formal objects that precisely represent these equivalence
classes of sequent proofs.  These objects are called proof nets, and
they have a strong geometric character, corresponding to additional
graph structure (``axiom links'') on the syntactical forest of the
sequent.  More precisely, given a sequent \(\Gamma =
A_{1},\ldots,A_{n}\) and a proof \(\pi\) of that sequent, then the
proof net that represents \(\pi\) is simply given by the syntactical
forest of \(\Gamma\) decorated with additional edges (shown in thick
lines in the picture below) that represent the identity axioms that
appeared in the proof:
$$
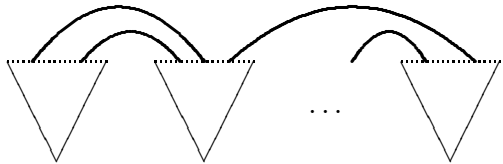
\begin{xy} 
0;<1.7em,0ex>:<0em,4ex>::
(0,1);(1,-1)**\dir{-},
(2,1);(1,-1)**\dir{-},
(0,1);(2,1) **\dir{.},
(3,1);(4,-1)**\dir{-},
(5,1);(4,-1)**\dir{-},
(3,1);(5,1) **\dir{.},
(8,1);(9,-1)**\dir{-},
(10,1);(9,-1)**\dir{-},
(8,1);(10,1) **\dir{.},
(6.5,0)*{\clap{\ldots}},
(0.5,1);(4,1)**\crv{~*=<.5pt>{.}(2.25,\doubleheight)},
(1.5,1);(3.5,1)**\crv{~*=<.5pt>{.}(2.5,\simpleheight)},
(4.5,1);(9.5,1)**\crv{~*=<.5pt>{.}(7,\doubleheight)},
(7,1);(8.5,1)**\crv{~*=<.5pt>{.}(7.75,\simpleheight)},
\end{xy}
$$
Moreover proof nets are vindicated by category theory, since the
category of two-formula sequents and proof nets is the free
*-autonomous category~\cite{barr:79} (if we omit units from the
definition of a *-autonomous category) on the set of generating atomic
formulas. This first appeared in~\cite{blute:93}, but the problem of
precisely defining a *-autonomous category without units has given
rise to recent
developements~\cite{lam:str:05:freebool,houston:etal:unitless,dosen:petric:05}.

As a matter of fact, axiom links were already visible, under the
name of \emph{Kelly-Mac Lane graphs} in the early
work~\cite{kelly:maclane:71} that tried to describe free symmetric
monoidal closed (also called \emph{autonomous}) categories; Girard's
key insights~\cite{girard:87} here were in noticing that there was an
inherent symmetry that could be formalized through a negation (thus
the move from autonomous to *-autonomous), and that the addition of
the axiom links to the sequent's syntactic forest were enough to
completely characterize the proof (if no units are present).

The theory of proof nets has been extended to larger fragments of
linear logic. When judged from the point of view of their ability to
identify proofs that should be identified, these extensions can be
shown to have varying degrees of success. One of these extensions,
which complies particularly well with the categorical ideal---and can
be firmly put in the ``successful'' class even without appealing to
categorical considerations---is the inclusion of additive connectives
presented in~\cite{hughes:glabbeek:03}, in which the additives
correspond exactly to categorical product and coproduct.

\bigskip In this paper we give a theory of proof nets for the full
multiplicative fragment, that is, including the multiplicative units.
We then prove that it allows us to construct the free *-autonomous
category with units on a given set of generating objects, thus getting
full validation from the categorical imperative.
 
When this paper was submitted there were only two other treatments of
multiplicative units that we were aware of. In~\cite{koh:ong:99}, the
authors provide an internal language for autonomous and *-autonomous
categories based on the $\lambda\mu$-calculus, and
in~\cite{blute:etal:96}, several classes of free monoidal categories
are constructed, by the means of a nonstandard version of two-sided
proof nets, with a correctness criterion which is a version of the
Danos contractibility criterion~\cite{danos:phd}.  Being based on the
\(\lambda\mu\)-calculus, the first of these papers is firmly in the
tradition of natural deduction, in which the logical rules
(introduction/elimination) mechanically generate the system of proof
objects, which are ordinary terms with binders. Unsurprisingly, an
equivalence relation on the terms is needed to construct the free
*-autonomous categories.  It is well-known that unless a calculus is
"intuitionistic" (with one-conclusion sequents), it is not easy at all
to establish a good correspondence between such systems of terms and
the graphical proof objects that have become the tradition in linear
logic; this is still research material.

As for the second of these papers, we think its approach is best
summarized, despite its title, by the means of the sequent
calculus. It starts with a core logic which is weaker than MLL, which
can be related to it as follows: Given the usual primitives $\ltens$,
$\lpar$, $\lone$, and $\lbot$ of multiplicative linear logic, look at
the following system of polarities, where \(\wht\) means ``right
side'' and \(\blk\) means ``left side'':
\[
\begin{array}{c}
  \begin{array}{c|cc}
    \ltens &\blk &\wht \\
    \hline 
    \blk  &\blk  &    \\
    \wht  &      & \wht
  \end{array}\qquad
  \begin{array}{c|cc}
    \lpar &\blk &\wht \\
    \hline 
    \blk  & \blk &    \\
    \wht  &      & \wht
  \end{array} \qquad 
  \begin{array}{c}
    \lone    \\
    \hline
    \blk   \\
    \wht  
  \end{array} \qquad
  \begin{array}{c}
    \lbot    \\
    \hline
    \blk   \\
    \wht  
  \end{array}
\end{array} 
\]
A constant may have either polarity, but you are only allowed to apply
a tensor or a par on two formulas that have the same polarity, and the
resulting formula has that same polarity. If we now add axioms of the
form \(a^{\wht},a^{\blk}\) then the main system
in~\cite{blute:etal:96} is exactly equivalent to multiplicative linear
logic with the usual rules, but where the introduction of connectives
have to obey the polarity restrictions above. A polarized one-sided
sequent of the form \(\seq
A_{1}^{\blk},\ldots,A_{n}^{\blk},B_{1}^{\wht},\ldots,B_{m}^{\wht}\) is
translated back in the authors' notation as
\(A_{n}\lneg,\ldots,A_{1}\lneg \seq B_{1},\ldots,B_{m}\), where the
\((-)\lneg\) operation here is the ordinary de~Morgan dual,
remembering that it inverts polarities. Thus it is indeed a weaker
logic than classical multiplicative linear logic, since, for instance,
ordinary axiom links always ``straddle the left-right divide''. But
there are no polarity restrictions on constants, and thus the
constant-only fragment, suitably quotiented, should give back the free
*-autonomous category generated by the empty set.  The addition of
non-logical axioms allows the construction of the free such category
generated by an ordinary category.

Later in~\cite{blute:etal:96}, a side-switching negation connective is
introduced, as is also done in~\cite{puite:phd}, along with
non-logical axioms for it, and the larger system is
equivalent to classical multiplicative linear logic.

In the next section we will say how our approach to proof nets differs
from the one which is used in~\cite{blute:etal:96}. It should be
obvious eventually that what we propose is considerably simpler. We
have chosen to use only sets of objects (atomic variables) as
generating sets. It would be easy to extend our work to construct the
free *-autonomous category generated by an arbitrary category, or an
arbitrary structad~\cite{lamarche:contexts}, but we are trying very
hard to be read by both algebraists and proof theorists, and proof
theorists are notoriously suspicious of non-logical axioms. In general
they kill cut-elimination/normalization, but there are indeed classes
of harmless non-logical axioms. Recently, general theoretical
criteria~\cite{dowek:werner:03} have been developed that allow the
identification of such harmless classes. 

Since this manuscript was submitted, yet another
approach~\cite{hughes:freestar} has been proposed to the problem
of constructing free *-autonomous categories, which improves on its
predecessors in that the cut-elimination process can be effected at
the level of the \emph{representatives} of the equivalence
classes~\cite{girard:96:PN}.

\subsection*{Outline of the paper}

This paper consists of two parts. The first one is only concerned with
syntax: the sequent calculus and the more modern syntax of proof nets,
and our variation on ordinary multiplicative proof nets that permits
the addition of units.  The standard results---sequentialization and
cut-elimination---are proved. The second part is concerned with
algebra: after some introductory material on autonomous and
*-autonomous categories, we show that, given a set \(\cA\) of atomic
formulas, the set of proof nets constructed in the previous section do
form a *-autonomous category, which is easy, and then that it is
actually the free one generated by \(\cA\), which is much harder.
Both sections have discussion on history and motivation. We have taken
pains to make the treatment as self contained as possible.  All the
proofs are complete; we have done the utmost to avoid any kind of
hand-waving, and we tried hard to ensure that a reader with only a
minimal background in multiplicative linear logic and/or category
theory could read this.

\medskip

The main results of this paper have already been presented at the
CSL-conference 2004 in Karpacz, but it was impossible to give the
complete story in fifteen pages \cite{str:lam:04:CSL}.


\section{Proof nets for multiplicative linear logic}\label{sec:pn}

We assume that the reader is familiar with the sequent calculus for
classical multiplicative linear logic, and has some basic notions on
proof nets, e.g., understands the idea of a correctness criterion.
For an introduction, the reader is referred to \cite{lafont:95,str:esslli06}.

\subsection{Why are the units problematic?}

The problem with the bottom rule is that it is very mobile. Suppose a
proof contains a rule instance $\rr$ which appears after a
\(\lbot\)-introduction rule and does not introduce a connective under
that \(\lbot\). Then $\rr$ can be pushed above that
\(\lbot\)-introduction:
$$
\vcenter{\ddernote{\rr}{}{\lbot,\Gamma'}
{\root{\lbot}{\lbot,\Gamma}{\leaf{\Gamma}}}{\leaf{\quad\cdots}}}
\qquad\longleftrightarrow\qquad
\vcenter{\dernote{\lbot}{}{\lbot,\Gamma'}
{\rroot{\rr}{\Gamma'}{\leaf{\enspace\Gamma\quad}}{\leaf{\quad\cdots}}}}
$$
It is very hard to find a good reason to decree that the difference
between these two sequent proofs has some essential significance,
which goes beyond mere notation, and that they ought to be
distinguished. Asking for a distinction opens the door to a theory of
proof identification which can only be slightly less bureaucratic than
the sequent calculus itself. Moreover, the theory of *-autonomous
categories tells us that they \emph{should} be identified.  But then
accepting this seemingly trivial permutation as an equation has deep
consequences.  Supposing that rule $\rr$ was a
\(\ltens\)-introduction, there is now a choice of two branches on
which to do the \(\lbot\)-introduction.
$$
\vcenter{\ddernote{\ltens}{}{\lbot,\Gamma,A\ltens B,\Delta}
{\root{\lbot}{\lbot,\Gamma,A}{\leaf{\Gamma,A}}}{\leaf{\quad B,\Delta}}}
\quad\longleftrightarrow\quad
\vcenter{\dernote{\lbot}{}{\lbot,\Gamma,A\ltens B,\Delta}
{\rroot{\ltens}{\Gamma,A\ltens
    B,\Delta}{\leaf{\Gamma,A\quad}}{\leaf{\quad B,\Delta}}}}
\quad\longleftrightarrow\quad
\vcenter{\ddernote{\ltens}{}{\lbot,\Gamma,A\ltens B,\Delta}
{\leaf{\Gamma,A\quad}}{\root{\lbot}{\lbot,B,\Delta}{\leaf{B,\Delta}}}}
$$

Ordinary proof nets for multiplicative linear logic have successfully
eliminated the bureaucracy introduced by the \(\ltens\) and \(\lpar\)
sequent rules. They are characterized by the presence of \emph{links},
which connect the atoms of the syntactical forest of the sequent. When
extending them to multiplicative units, the first impulse is probably
to try to attach the~\(\lbot\)s that are present on the sequent forest
on other atomic formulas. This corresponds to doing the
\(\lbot\)-introductions as early as possible, that is, as high up on
the sequent tree as can be done. This approach has very recently been
used in a most satisfying
manner~\cite{hughes:simple-mult,hughes:freestar}.

We see that an arbitrary choice has to be made because of tensor
introductions: in a \(\ltens\)-introduction one of the two branches of
the sequent proof tree has to be chosen for doing the
\(\lbot\)-introduction. In such a situation, correct identification of
proofs can only be achieved by considering equivalences classes of
graphs, and the theory of proof nets involves an equivalence relation
on a set of ``correct'' graphs.

\begin{figure}[t]
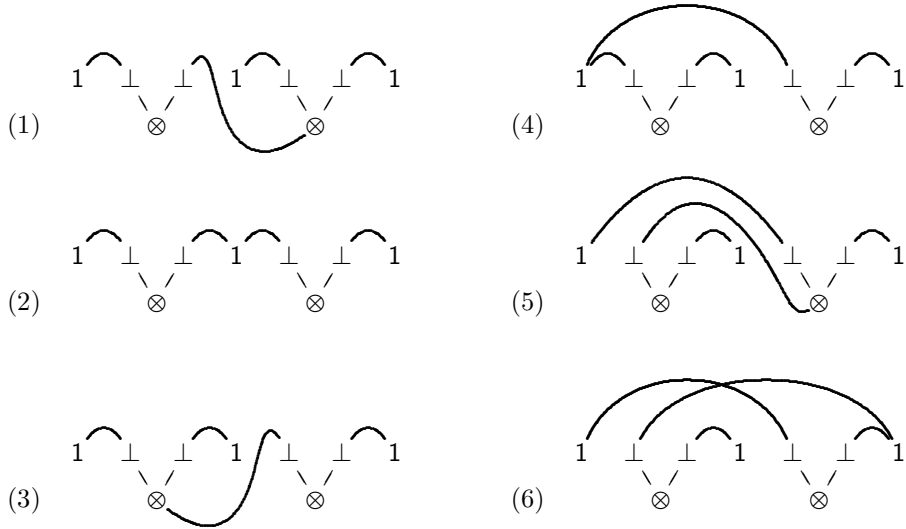

\small
\vskip-.5ex
$$
(1)\quad
\begin{introdia}
  "a"*{\pbot};"b"*{\pbot} **\crv{~*=<.5pt>{.}(1,\simpleheight)},
  "d"*{\pbot};"e"*{\pbot} **\crv{~*=<.5pt>{.}(7,\simpleheight)},
  "f"*{\pbot};"g"*{\pbot} **\crv{~*=<.5pt>{.}(11,\simpleheight)},
  "c"*{\pbot};"i"*{\pbot} 
  **\crv{~*=<.5pt>{.}(5,\simpleheight)&(5,0)&(7,\lowerheight)},  
\end{introdia}
\qquad\qquad
(4)\quad
\begin{introdia}
  "a"*{\pbot};"b"*{\pbot} **\crv{~*=<.5pt>{.}(1,\simpleheight)},
  "c"*{\pbot};"d"*{\pbot} **\crv{~*=<.5pt>{.}(5,\simpleheight)},
  "f"*{\pbot};"g"*{\pbot} **\crv{~*=<.5pt>{.}(11,\simpleheight)},
  "a"*{\pbot};"e"*{\pbot} 
  **\crv{~*=<.5pt>{.}(1,\doubleheight)&(7,\doubleheight)},
\end{introdia}
$$
\vskip-1ex
$$
(2)\quad
\begin{introdia}
  "a"*{\pbot};"b"*{\pbot} **\crv{~*=<.5pt>{.}(1,\simpleheight)},
  "c"*{\pbot};"d"*{\pbot} **\crv{~*=<.5pt>{.}(5,\simpleheight)},
  "d"*{\pbot};"e"*{\pbot} **\crv{~*=<.5pt>{.}(7,\simpleheight)},
  "f"*{\pbot};"g"*{\pbot} **\crv{~*=<.5pt>{.}(11,\simpleheight)},
\end{introdia}
\qquad\qquad
(5)\quad
\begin{introdia}
  "a"*{\pbot};"e"*{\pbot} **\crv{~*=<.5pt>{.}(4,\tripleheight)},
  "c"*{\pbot};"d"*{\pbot} **\crv{~*=<.5pt>{.}(5,\simpleheight)},
  "f"*{\pbot};"g"*{\pbot} **\crv{~*=<.5pt>{.}(11,\simpleheight)},
  "b"*{\pbot};"i"*{\pbot} 
  **\crv{~*=<.5pt>{.}(4,\doubleheight)&(8,0.5)&(8,\lowerhalf)},  
\end{introdia}
$$
\vskip-1ex
$$
(3)\quad
\begin{introdia}
  "a"*{\pbot};"b"*{\pbot} **\crv{~*=<.5pt>{.}(1,\simpleheight)},
  "c"*{\pbot};"d"*{\pbot} **\crv{~*=<.5pt>{.}(5,\simpleheight)},
  "f"*{\pbot};"g"*{\pbot} **\crv{~*=<.5pt>{.}(11,\simpleheight)},
  "e"*{\pbot};"h"*{\pbot} 
  **\crv{~*=<.5pt>{.}(7,\simpleheight)&(7,0)&(5,\lowerheight)},  
\end{introdia}
\qquad\qquad
(6)\quad
\begin{introdia}
  "c"*{\pbot};"d"*{\pbot} **\crv{~*=<.5pt>{.}(5,\simpleheight)},
  "f"*{\pbot};"g"*{\pbot} **\crv{~*=<.5pt>{.}(11,\simpleheight)},
  "a"*{\pbot};"e"*{\pbot} 
  **\crv{~*=<.5pt>{.}(1,\doubleheight)&(7,\doubleheight)},
  "b"*{\pbot};"g"*{\pbot} 
  **\crv{~*=<.5pt>{.}(3,\doubleheight)&(11,\doubleheight)},
\end{introdia}
$$
\caption{Different representations of the same proof}
\label{fig:proofgraphs}
\end{figure}
Another possibility is to attach the \(\lbot\)s on \emph{branches} of
the proof forest. This is done in~\cite{blute:etal:96}, where an
equivalence class is also used, built on the ability to slide the
constants up and down the branches.

Yet another possibility is to attach the \(\lbot\)s ``as low as
possible'' on the forest, corresponding to the idea that in the
sequent calculus deduction the \(\lbot\)-introduction would be done as
late as possible, for example just before the \(\lbot\) instance gets
a connective introduced under it. One way of implementing this is
linking the \(\lbot\) instance to the last connective that was
introduced above it. This is not the only way of doing things, for
example we could imagine links that attach that \(\lbot\) instance to
several subformulas of the sequent forest, corresponding to the
several conclusions of the sequent that existed above the
\(\lbot\)-introduction.\footnote{Our approach is probably best seen as
  a version of this, where ordinary proof net technology is used to
  express this idea. We add extra ``bunching'' nodes, that act like
  ordinary pars from the point of view of correctness.}
But whatever way we choose to ``normalize'' proofs, we claim that if
the conventional notion of ``link'' is used for
\(\lbot\)s,\footnote{I.e., if we consider a proof \(\pi\) on the
  sequent \(\Gamma\) as the sequent forest of \(\Gamma\) decorated
  with special edges that encode information about the essence of
  \(\pi\).} we still need to use equivalence classes of such graphs,
and there is no hope of having a normal form in that universe of
enriched sequent graphs. For instance, the six graphs in
Figure~\ref{fig:proofgraphs} are easily seen to represent equivalent
proofs, because going from an odd-numbered example to its successor is
just sliding a \(\lbot\)-intro up in one of the \(\ltens\)-intro
branches, and going from an even-numbered example to its successor is
just doing the reverse transformation. But notice that examples
\((3)\) and \((5)\) are \emph{distinct but isomorphic} graphs, since
one can be exactly superposed on the other \emph{by only using the
  Exchange rule}. Thus it is impossible, given the information at our
disposal, to choose one instead of the other to represent the abstract
proof they both denote.  The only way this could be done would be by
using arbitrary extra information, like the order of the formulas in
the sequent, a strategy that only replaces the overdeterminism of the
sequent calculus by another kind of overdeterminism.

The same can be said of Examples \((2)\) and \((6)\), which are also
isomorphic modulo Exchange. But notice that these two comply to the
``as early as possible'' strategy, while the previous two were of the
``as late as possible'' kind. So for neither strategy can there be a
hope a graphical normal form. The interested reader can verify that
the six examples above are part of a ``ring'' of 24 graphs that are
all equivalent from the point of view of category theory. Thus there
are essentially only two proofs of that sequent, but 48 possible
different graphs like these on it.

Thus there is one aspect of our work that does not differ
from~\cite{blute:etal:96}, which is our presentation of abstract
proofs as equivalence classes of graphs. But some related aspects are
significantly different:
\begin{itemize}
\item The graphs that belong to our equivalence classes are
  \emph{standard multiplicative proof nets,} where the usual notions,
  like correctness criteria and the empire of a tensor branch, will
  apply.  It is just that some \(\lpar\) and \(\ltens\) links are used
  in a particular fashion to deal with the units.  (The readers can
  choose their favorite correctness criterion since they are all
  equivalent; in this paper we will use the one of
  \cite{danos:regnier:89} because of its popularity.)
\item The equivalence relation we will present is based on a very simple
  set of rewriting rules on proof graphs. As a matter of fact, there
  is only \emph{one} non-trivial rule, since the other rules have to
  do with commutativity and associativity of the connectives and can be
  dispensed with if we use, for example, \(n\)-ary connectives.
\end{itemize}

\subsection{Cut free proof nets}\label{sec:nets}

Let $\cA=\set{a,b,\dots}$ be an arbitrary set of atoms, and let
$\cA\lneg=\set{a\lneg,b\lneg,\dots}$. The set of MLL \dfn{formulas} is
defined as follows:
\begin{equation}
  \label{eq:formula}
  \cF \grammareq \cA \mid \cA\lneg \mid \lone \mid \lbot \mid
  \cF\ltens\cF \mid \cF\lpar\cF
  \quadfs
\end{equation}
Additionally, we will define the set of MLL \dfn{linkings} (which can
be seen as a special kind of formulas) as follows:
\begin{equation}
  \label{eq:linking}
  \cL \grammareq \lone \mid a\ltens a\lneg\mid a\lneg\ltens a 
  \mid \lbot\ltens\cL
  \mid \cL\ltens\lbot \mid \cL\lpar\cL\quadfs
\end{equation}
Here, $a$ stands for any element of $\cA$.  We will use $A$, $B$,
\ldots\ to denote formulas, and $P$, $Q$, \ldots\ to denote linkings.
\dfn{Sequents} (denoted by $\Gamma$, $\Delta$, \ldots) are finite
lists of formulas (separated by comma).

In the following, we will consider formulas and linkings always as
binary trees (and sequents as forests), whose leaves are decorated by
elements of $\cA\cup\cA\lneg\cup\set{\lone,\lbot}$, and whose inner
nodes are decorated by $\lpar$ or $\ltens$. We can also think of the
nodes being decorated by the whole subformula rooted at that node.

\begin{figure}[t]
$$
\begin{xy} 
0;<1.7em,0ex>:<0em,4ex>::
(0,1);(1,-1)**\dir{-},
(2,1);(1,-1)**\dir{-},
(0,1);(8,1) **\dir{.},
(2,1);(3,-1)**\dir{-},
(4,1);(3,-1)**\dir{-},
(6,1);(7,-1)**\dir{-},
(8,1);(7,-1)**\dir{-},
(5,0)*{\clap{\ldots}},
(0,1);(4,3)**\dir{-},
(8,1);(4,3)**\dir{-},
(8.5,2.25)*{\rlap{$\leftarrow$ linking tree}},
(8.5,1)*{\rlap{$\leftarrow$ leaves}},
(8.5,-0.25)*{\rlap{$\leftarrow$ sequent forest}},
\end{xy}
$$
\caption{Linking tree and sequent forest sharing the same set of leaves}
\label{fig:proofgraph}  
\end{figure}
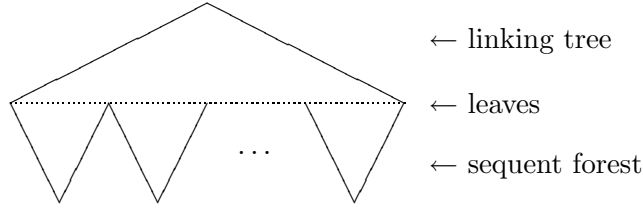

\begin{definition}
  A \dfn{pre-proof graph} is a graph consisting of a linking $P$ and a
  sequent~$\Gamma$, such that the set of leaves of \(P\) coincides
  with the set of leaves of \(\Gamma\) (as depicted in
  Figure~\ref{fig:proofgraph}).  It will be denoted by
  $\prfgrf{P}{\Gamma}$ or by $$\prfgrf{P}{\Gamma}\quadfs$$
\end{definition}
 
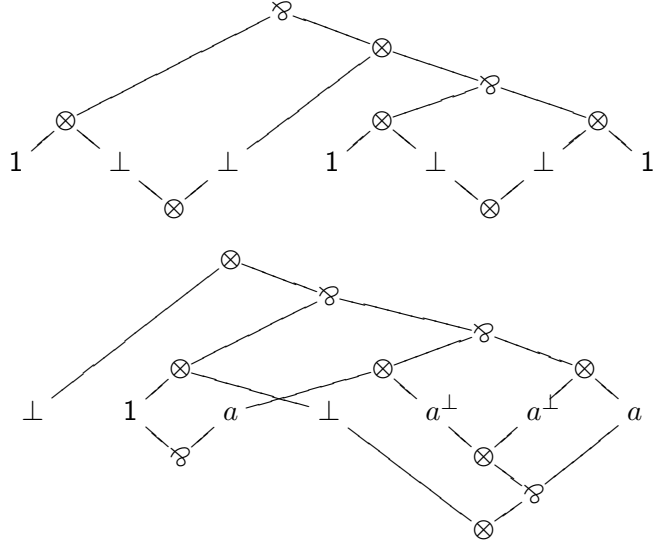
\begin{figure}[t]
\vskip-1ex
$$
\xymatrix@C=\matrcol @R=\matrrow{\ar@{-}
&&&&&\mpar{dddllll}{drr}\\
&              &&&&&&\mtens{dddlll}{drr}\\
&              &&&&&&&&\mpar{dll}{drr}\\
&\mtens{dl}{dr}&&&&&&\mtens{dl}{dr}&&&&\mtens{dl}{dr}\\
\mone&&\mbot&&\mbot&&\mone&&\mbot&&\mbot&&\mone\\
&&&\mtens{ul}{ur}&&&&&&\mtens{ul}{ur}\\
}
$$
$$
\xymatrix@C=\matrcol @R=\matrrow{\ar@{-}
&&&&\mtens{ddddllll}{drr}\\
&&&                &&                &\mpar{ddlll}{drrr}\\
&&&                && &&&&\mpar{dll}{drr}\\
&&&\mtens{dl}{drrr}&&&&\mtens{dlll}{dr}&&&&\mtens{dl}{dr}\\
\mbot&&\mone&&\mat{a}&&\mbot&&\matn{a}&&\matn{a}&&\mat{a}\\
&&&\mpar{ul}{ur}&&&&&&\mtens{ul}{ur}\\
&&&          &&&&&&&\mpar{ul}{uurr}\\
&&&             &&&&&&\mtens{uuulll}{ur}\\
}
$$
\vskip-1ex
\caption{Two examples of proof graphs}
\label{fig:grfexa}
\end{figure}

Following the tradition, we will draw these graphs in such a way that
the roots of the formula trees are at the bottom, the root of the
linking tree is at the top, and the leaves are in between.
Figure~\ref{fig:grfexa} shows two examples. The first of them
corresponds to the first graph in Figure~\ref{fig:proofgraphs}.  In a
more compact notation we will write them as
\begin{equation}
  \label{eq:exa1}
\prfgrf
{(\lone_1\ltens\lbot_2)\lpar(\lbot_3\ltens((\lone_4\ltens\lbot_5)\lpar
(\lbot_6\ltens\lone_7)))}
{\lone_1,\lbot_2\ltens\lbot_3,\lone_4,\lbot_5\ltens\lbot_6,\lone_7}
\end{equation}
and
\begin{equation}
  \label{eq:exa2}
\prfgrf%
{\clap{$\lbot_1\ltens((\lone_2\ltens\lbot_4)\lpar
((a_3\ltens a\lneg_5)\lpar(a_6\lneg\ltens a_7)))$}}%
{\lbot_1,\lone_2\lpar a_3,
\lbot_4\ltens((a\lneg_5\ltens a\lneg_6)\lpar a_7)}
\end{equation}
Here, the indices are used to show how the leaves of the linking and
the leaves of the sequent are identified. In this way we will,
throughout this paper, use indices on atoms 
to distinguish between different
occurrences of the same atom (\ie $a_3$ and $a_7$ do not denote
different elements of $\cA$). 
In the same way, indices on the units
$\lone$ and $\lbot$ are used to distinguish different occurrences.

\begin{remark}
  Since we are dealing with sets-and-structure, we should mention an
  additional bit of structure that pre-proof-graphs possess, which is
  a total ordering on the set of leaves, corresponding to the
  syntactic order in which formulas and sequents are written. Since
  this order is completely obvious in the notation, we will
  not mention it again. Thus we take the most standard approach to the
  sequent calculus in commutative logic, in which a sequent is a
  \emph{sequence} of formulas, subject to the permutations that are
  induced by the Exchange rule.
\end{remark}

\begin{definition}\label{def:switching}
  A \dfn{switching} of a pre-proof graph $\prfgrf{P}{\Gamma}$ is a
  graph $G$
  that is obtained from $\prfgrf{P}{\Gamma}$ by omitting for each
  $\lpar$-node one of the two edges that connect the node to its
  children.
  \cite{danos:regnier:89}
\end{definition}

\begin{definition}\label{def:correct}
  A pre-proof graph $\prfgrf{P}{\Gamma}$ is called \dfn{correct} if all its
  switchings are connected and acyclic.
  A \dfn{proof graph} is a correct pre-proof graph.
\end{definition}

The examples in Figure~\ref{fig:grfexa} are proof graphs.

Let $\prfgrf{P}{\Gamma}$ be a pre-proof graph where one $\lbot$ is
selected. Let it be indexed as $\lbot_i$. Now, let $G$ be a switching
of $\prfgrf{P}{\Gamma}$, and let $G'$ be the graph obtained from $G$
by removing the edge between $\lbot_i$ and its parent in $P$ (which is
always a $\ltens$). Then $G'$ is called an \dfn{extended switching} of
$\prfgrf{P}{\Gamma}$ with respect to $\lbot_i$. Observe that, if
$\prfgrf{P}{\Gamma}$ is correct, then every extended switching is
disconnected and consists of two connected components 
(see \cite{fleury:retore:94} for a discussion on connected components
in switchings).

We will use the notation $\prfgrf{P\cons{Q}}{\Gamma}$ to distinguish
the subtree $Q$ of the linking tree of the graph. Then $P\conhole$ is
the context of $Q$.

\begin{para}{\bf Equivalence on pre-proof graphs.}\label{par:grfequ}
On the set of pre-proof graphs we will define the relation $\prfequ$ to
be the smallest equivalence relation satisfying 
$$
\begin{array}{r@{\;\;}c@{\;\;}l}
  \prfgrf{P\cons{Q\lpar R}}{\Gamma} &\prfequ& 
  \prfgrf{P\cons{R\lpar Q}}{\Gamma}\\
  \prfgrf{P\cons{(Q\lpar R)\lpar S}}{\Gamma} &\prfequ& 
  \prfgrf{P\cons{Q\lpar (R\lpar S)}}{\Gamma}\\
  \prfgrf{P\cons{Q\ltens R}}{\Gamma} &\prfequ& 
  \prfgrf{P\cons{R\ltens Q}}{\Gamma}\\
  \prfgrf{P\cons{\lbot_i\ltens(Q\ltens\lbot_j)}}{\Gamma} &\prfequ& 
  \prfgrf{P\cons{(\lbot_i\ltens Q)\ltens\lbot_j}}{\Gamma}\\[.7ex]
  \prfgrf{P\cons{Q\lpar(R\ltens\lbot_i)}}{\Gamma} &\stackrel{(*)}{\prfequ}& 
  \prfgrf{P\cons{(Q\lpar R)\ltens\lbot_i}}{\Gamma}\quadcm\quad
\end{array}
$$
where the last equation only holds if the following side condition is
fulfilled:
{
\def\labelitemi{$(*)$}
\begin{itemize}
\item In every extended switching of 
$\prfgrf{P\cons{Q\lpar(R\ltens\lbot_i)}}{\Gamma}$ with respect to
$\lbot_i$ no node of the subtree $Q$ is connected to $\lbot_i$.
\end{itemize}
}
In all equations $Q$, $R$ and $S$ are arbitrary linking trees and
$P\conhole$ is an arbitrary linking tree context. $\Gamma$ is an
arbitrary sequent such that its leaves match those of the
linking.
\end{para}

The following proof graph is equivalent to the second one in
Figure~\ref{fig:grfexa}, i.e., to \eqref{eq:exa2}:
$$
\prfgrf%
{\clap{$(((\lbot_1\ltens\lone_2)\ltens\lbot_4)\lpar
(a_3\ltens a\lneg_5))\lpar(a\lneg_6\ltens a_7)$}}%
{\lbot_1,\lone_2\lpar a_3,
\lbot_4\ltens((a\lneg_5\ltens a\lneg_6)\lpar a_7)}
$$

\begin{definition}\label{def:proofnet}
  A \dfn{pre-proof net}\footnote{What we call \dfn{pre-proof net} is
  in the literature often called \dfn{proof structure.}} 
  is an equivalence class
  $\prfneteq{P}{\Gamma}$. A pre-proof net is \dfn{correct} if one of its
  elements is correct. In this case it is called a \dfn{proof net}.
\end{definition}

In the following, we will for a given proof graph $\prfgrf{P}{\Gamma}$
write $\prfnet{P}{\Gamma}$ to denote the proof net formed by its
equivalence class (\ie we will omit the $\prfequ$~subscript).
 
\begin{lemma}\label{lem:equcorrect}
  If $\prfgrf{P}{\Gamma}$ is correct and
  $\prfgrf{P}{\Gamma}\prfequ\prfgrf{P'}{\Gamma}$, then
  $\prfgrf{P'}{\Gamma}$ is also correct.
\end{lemma}

\proof
That the first four equations preserve correctness is obvious. If in
the last equation there is a switching that makes one sided disconnected, then
it also makes the other side disconnected. For acyclicity, we have to check
whether there is a switching that produces a cycle on the right-hand
side of the equation and not on the left-hand side. This is only
possible if the cycle contains some nodes of $Q$ and the
$\lbot_i$. But this case is ruled out by the side condition $(*)$.
\qed 

Lemma~\ref{lem:equcorrect} ensures that the notion of proof net is
well-defined, in the sense that all its members are proof graphs, \ie
correct. 

An alternative approach to the definition of a proof net would be to
restrict the definition of \(\prfequ\) to proof graphs and not mention
pre-proof graphs at all, i.e., to assume from the start that everything
obeys the correctness criterion. This slight breach with tradition
has the advantage of not requiring the side condition, which is
asymmetrical. Such a point of view is used systematically
in~\cite{hughes:freestar}.

\subsection{Sequentialization}\label{sec:sequent}

In this section we will relate our proof nets to cut free sequent calculus
proofs of MLL. For this, we will first show, how cut free sequent proofs of
MLL can be inductively translated into pre-proof graphs. This is done in
Figure~\ref{fig:seq-pn}.
We will call a pre-proof net \dfn{sequentializable} if one of its
representatives can be obtained from a sequent calculus proof via this
translation. 


\begin{theorem}\label{thm:sequent}
  A pre-proof net is sequentializable if and only if it is a proof net.
\end{theorem}

\begin{figure}[t]
  $$
  \begin{array}{c@{\qquad\qquad}c}
  \vcinf{\idr}{\prfgrf{a\ltens a\lneg}{a,a\lneg}}{}&
  \vcinf{\exr}{\prfgrf{P}{\Gamma,B,A,\Delta}}{\prfgrf{P}{\Gamma,A,B,\Delta}}
  \\[4ex]
  \vcinf{\lone}{\prfgrf{\lone}{\lone}}{}
  &
  \vcinf{\lbot}{\prfgrf{\lbot\ltens P}{\lbot,\Gamma}}{\prfgrf{P}{\Gamma}}
  \\[4ex]
  \vcinf{\lpar}{\prfgrf{P}{A\lpar B,\Gamma}}{\prfgrf{P}{A,B,\Gamma}}
  &
  \vciinf{\ltens}{\prfgrf{P\lpar Q}{\Gamma,A\ltens B,\Delta}}
         {\prfgrf{P}{\Gamma,A}\;}{\;\prfgrf{Q}{B,\Delta}}
  \end{array}
  $$
   \caption{Translation of cut free 
     sequent calculus proofs into pre-proof
     graphs}
   \label{fig:seq-pn}
\end{figure}

For the proof we will need the observation that any proof graph is an
ordinary proof net (in the sense of \cite{danos:regnier:89}), 
and the well-known fact that there is
always a splitting tensor in such a net.

\begin{para}{\bf Ordinary proof nets.}\label{par:unitfree}
  An \dfn{ordinary axiom linking} for a sequent $\Gamma$ is a perfect
  matching of the leaves of $\Gamma$ (\ie the atoms and units) such
  that only dual atoms or units are matched. Of course, there are
  sequents for which no ordinary axiom linking exists, for example
  $a,b\lneg\lpar c,a\lneg\ltens c$, and there are sequents with more
  than one possible ordinary axiom linking, for example
  $a\lpar a,a\lneg\ltens a\lneg$. An \dfn{ordinary pre-proof net} is a
  sequent $\Gamma$ equipped with an ordinary axiom linking. Switchings
  and correctness are defined as in Definitions~\ref{def:switching}
  and~\ref{def:correct}. An \dfn{ordinary proof net} is a correct
  ordinary pre-proof net. In other words, what we call ordinary proof
  nets are the nets that are thoroughly studied in the literature,
  with the only difference that we also allow $\lone$ and~$\lbot$ at
  the places of the leaves.
\end{para}

\begin{figure}
  $$
  \begin{xy}
    0;<1.4em,0ex>:<0em,4ex>::
    (0,1)="a"*{\sna}, (2,1)="b"*{\sa}, (4,1)="c"*{\sa},
    (6,1)="d"*{\sna}, (8,1)="e"*{\sone}, (10,1)="f"*{\sbot},
    (12,1)="g"*{\sone}, (14,1)="gg"*{\sbot},
    (16,1)="ff"*{\sone}, (18,1)="dd"*{\sa}, (20,1)="ee"*{\sbot},
    (22,1)="cc"*{\sna}, (24,1)="bb"*{\sna}, (26,1)="aa"*{\sa},
    (1,0)="h"*{\ltens}, (5,0)="i"*{\ltens},
    (3,-1)="j"*{\lpar}, (9,0)="k"*{\ltens},
    (5.5,-2)="l"*{\lpar}, (7,-3)="m"*{\ltens},
    (17,0)="n"*{\lpar}, (23,0)="o"*{\ltens},
    (24,-1)="p"*{\lpar}, (23,-2)="q"*{\ltens},
    "a"*{\ptens};"h"*{\ptens} **\dir{-},
    "b"*{\ptens};"h"*{\ptens} **\dir{-},
    "c"*{\ptens};"i"*{\ptens} **\dir{-},
    "d"*{\ptens};"i"*{\ptens} **\dir{-},
    "e"*{\ptens};"k"*{\ptens} **\dir{-},
    "f"*{\ptens};"k"*{\ptens} **\dir{-},
    "i"*{\ptens};"j"*{\ptens} **\dir{-},
    "h"*{\ptens};"j"*{\ptens} **\dir{-},
    "j"*{\ptens};"l"*{\ptens} **\dir{-},
    "k"*{\ptens};"l"*{\ptens} **\dir{-},
    "g"*{\ptens};"m"*{\ptens} **\dir{-},
    "l"*{\ptens};"m"*{\ptens} **\dir{-},
    "ff"*{\ptens};"n"*{\ptens} **\dir{-},
    "dd"*{\ptens};"n"*{\ptens} **\dir{-},
    "cc"*{\ptens};"o"*{\ptens} **\dir{-},
    "bb"*{\ptens};"o"*{\ptens} **\dir{-},
    "aa"*{\ptens};"p"*{\ptens} **\dir{-},
    "o"*{\ptens};"p"*{\ptens} **\dir{-},
    "p"*{\ptens};"q"*{\ptens} **\dir{-},
    "ee"*{\ptens};"q"*{\ptens} **\dir{-},
    "g"*{\ptens};"gg"*{\ptens} **\crv{~*=<.5pt>{.}(13,\simpleheight)},
    "f"*{\ptens};"ff"*{\ptens} **\crv{~*=<.5pt>{.}(13,\doubleheight)},
    "e"*{\ptens};"ee"*{\ptens} **\crv{~*=<.5pt>{.}(14,\tripleheight)},
    "d"*{\ptens};"dd"*{\ptens} **\crv{~*=<.5pt>{.}(12,\tripleheight)},
    "c"*{\ptens};"cc"*{\ptens} **\crv{~*=<.5pt>{.}(13,\quadrupleheight)},
    "b"*{\ptens};"bb"*{\ptens} **\crv{~*=<.5pt>{.}(13,\quintupleheight)},
    "a"*{\ptens};"aa"*{\ptens} **\crv{~*=<.5pt>{.}(13,\sextupleheight)},
  \end{xy}
  $$
  \caption{Example of an ordinary proof net}
  \label{fig:ordexa}
\end{figure}
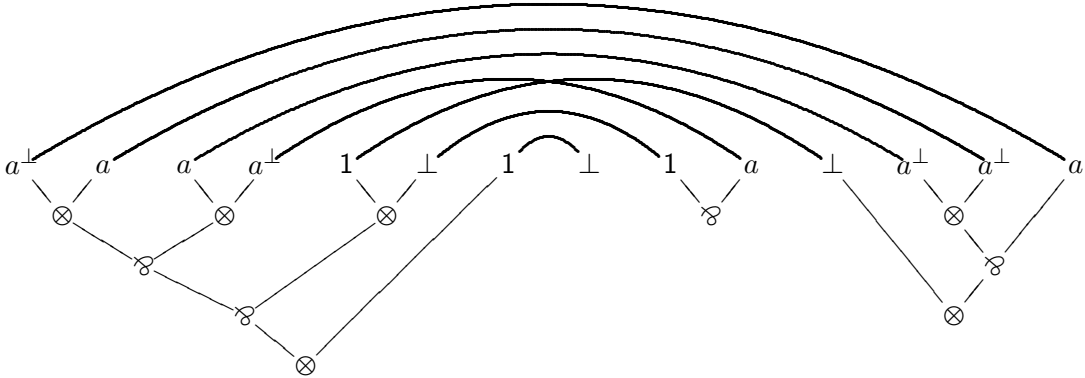

\begin{observation}\label{obs:unitfree}
  Every pre-proof graph $\prfgrf{P}{\Gamma}$ is an ordinary
  pre-proof net. To
  make this precise, define for the linking $P$ the \dfn{linking formula}
  $P\lfmark$ inductively as follows:
  $$
  \begin{array}{r@{\;=\;}l@{\quad}r@{\;=\;}l@{\qquad}r@{\;=\;}l}
    a\lneglfmark&a&\lone\lfmark&\lbot&(A\ltens B)\lfmark&
    B\lfmark\ltens A\lfmark\\
    a\lfmark&a\lneg&\lbot\lfmark&\lone&(A\lpar B)\lfmark&
    B\lfmark\lpar A\lfmark
  \end{array}
  \quad.
  $$
  In other words, $P\lfmark$ is obtained from $P$ by first
  replacing each leaf by its dual and by leaving all inner nodes
  unchanged, and then taking the mirror image of the
  tree\footnote{Since we are dealing only with the commutative case,
    taking the mirror image is unnecessary. However it simplifies many
    of the diagrams, avoiding unnecessary crossings, and this as much
    for proof nets as for commutative diagrams.}.  We now connect the
  leaves of $P\lfmark$ and $\Gamma$ by ordinary axiom links according
  to the leaf identification in $\prfgrf{P}{\Gamma}$.  Here we forget
  the fact that $\lbot$ and $\lone$ are the units and think of them as
  ordinary dual atoms. We get an ordinary pre-proof net, which we will
  denote by $\unitfree{P}{\Gamma}$.  Figure~\ref{fig:ordexa} shows as
  example the ordinary proof net obtained from the second proof graph
  in
  Figure~\ref{fig:grfexa}.%
  \footnote{If $\Gamma$ consists of only one formula, then we have an
  object which is in \cite{balat:dicosmo:99} called a \emph{bipartite
  proof net}. In fact, two proof graphs (in our sense) are equivalent
  if and only if the two linkings (seen as formulas) are isomorphic
  (in the sense of \cite{balat:dicosmo:99}).}
  Observe that $\unitfree{P}{\Gamma}$
  is correct if and only if $\prfgrf{P}{\Gamma}$ is correct.
\end{observation}

\begin{lemma}\label{lem:splitting}
  If in a ordinary proof net all roots are $\ltens$-nodes, then one
  of them is splitting, \ie by removing it the net becomes
  disconnected. \rm\cite{girard:87}
\end{lemma}

There are several proofs available for this lemma---one example is
Girard's original paper \cite{girard:87}, and another (very
instructive) paper is Retor\'e's \cite{retore:03}. For this reason we
do not show the proof here and concentrate on the

\proof[Proof of Theorem \ref{thm:sequent}]
  It is easy to see that the rules $\lone$ and $\idr$ give proof
  graphs and that the rules $\lbot$, $\lpar$, and $\ltens$ preserve
  the correctness. Therefore every sequentializable pre-proof net is 
  correct.

  For the other direction 
  pick one representative $\prfgrf{P}{\Gamma}$ of the proof
  net and proceed by induction on the sum of the number of $\ltens$-nodes 
  in the graph and the number of $\lpar$-nodes in $\Gamma$. In other
  words, the number of $\lpar$-nodes in $P$ is not relevant.
  (We will end up by exhibiting a sequentialization of an equivalent
  proof graph $\prfgrf{Q}{\Gamma}$, obtained from $\prfgrf{P}{\Gamma}$
  by only applying associativity and commutativity of $\lpar$, \ie the first
  two equations in \ref{par:grfequ}.)

  The base case is trivial (the graph consists of a single node which
  is labeled by~$\lone$). For the inductive case look at the root
  nodes in  $\Gamma$. If one of them is a $\lpar$, we can
  remove it by applying the $\lpar$-rule and
  proceed by induction hypothesis. If all roots in $\Gamma$
  are $\ltens$ nodes, we interpret $\prfgrf{P}{\Gamma}$ as an ordinary
  ordinary proof net (according to Observation~\ref{obs:unitfree}),
  which remains correct if we remove in $P$ all $\lpar$-nodes that
  do not have a $\ltens$-node as ancestor. Now all roots are
  $\ltens$-nodes and one of them is splitting 
  (by Lemma~\ref{lem:splitting}). If it belongs to $\Gamma$, we
  restore the $\lpar$-structure above the $\ltens$-nodes in $P$ such
  that each of the two subtrees of the root-$\lpar$ covers one of the
  two components in which the graph is divided by removing the
  splitting $\ltens$. We can now apply the $\ltens$-rule and proceed by
  induction hypothesis on the two premises. If the
  splitting $\ltens$ belongs to $P$, there
  are two possibilities. Either it comes from 
  an axiom link (\ie both children
  are dual atoms), or it comes from a bottom link (\ie one child is a
  $\lbot$).
  In the first case, we have that the graph is of the shape
  $\prfgrf{P',a_i\ltens
  a_j\lneg,P''}{\Gamma'\cons{a_i},\Gamma''\cons{a_j\lneg}}$, where the
  linking is written as sequent because the $\lpar$-roots are removed,
  and where $\Gamma'\cons{a_i}$ denotes a sequent where one formula contains
  the atom $a$, indexed as $a_i$, such that $P',a_i$ and
  $\Gamma'\cons{a_i}$ share the same atoms and units, 
  as well as $a_j\lneg,P''$
  and $\Gamma''\cons{a_j\lneg}$. If we replace $a_i$ by
  $\lone_i$ and $a_j\lneg$ by $\lone_j$, we obtain two proof graphs
  $\prfgrf{P',\lone_i}{\Gamma'\cons{\lone_i}}$ and 
  $\prfgrf{\lone_j,P''}{\Gamma''\cons{\lone_j}}$ of strictly smaller size.
  Therefore, by induction hypothesis, we have two sequent proofs
  \proofadjust
  \def\thisderisizea{1.6}
  \def\thisderisizeb{1.3}
  $$
  \DerivationFactors{\quad}%
                    {\quad\quad\quad}%
                    {\qlapm{\vcinf{\lone}{\;\lone_i}{}\quad}}%
                    {\Gamma'\cons{\lone_i}}%
                    {\pi_1}{\thisderisizea}{\thisderisizeb}
  \qquad\mbox{and}\qquad
  \DerivationFactors{\qlapm{\vcinf{\lone}{\;\lone_j}{}\quad}}%
                    {\quad\quad\quad}%
                    {\quad}%
                    {\Gamma''\cons{\lone_j}}%
                    {\pi_2}{\thisderisizea}{\thisderisizeb}
                    \quadfs
                    $$
  From $\pi_1$ and $\pi_2$ we can construct the following sequent
  proof:
  $$
  \DerivationFactors
      {\quad}%
      {\quad\quad\quad}%
      {\qlapm{\DerivationFactors
        {\qlapm{\vcinf{\idr}{a_i,a_j\lneg}{}}}%
        {\quad\quad\quad}%
        {\quad}%
        {a_i,\Gamma''\cons{a_j\lneg}}%
        {\pi'_2}{\thisderisizea}{\thisderisizeb}}}%
      {\Gamma'\cons{a_i},\Gamma''\cons{a_j\lneg}}%
      {\pi'_1}{\thisderisizea}{\thisderisizeb}
      \quadcm
      $$
  where $\pi'_1$ is obtained from $\pi_1$ by replacing
  $\lone_i$ everywhere by $a_i$ and by adding $\Gamma''\cons{a_j\lneg}$
  everywhere to the sequent that contains the $a_i$. Similarly
  $\pi'_2$ is obtained from $\pi_2$. It is easy to see that this proof
  translates into $\prfnet{P}{\Gamma}$.
  The case where the splitting tensor in $P$ has a $\lbot$ as child is
  similar and left to the reader.
  \qed

\begin{remark}
  It is well known that ordinary proof nets also have a
  sequentialization theorem, \ie they are correct if and only they are
  obtained from a (unit-free) sequent calculus proof in the obvious
  way. This has been studied thoroughly in the literature (\eg
  \cite{girard:87,danos:regnier:89,retore:03}).
\end{remark}

\subsection{Proof nets with cuts}\label{sec:cut}

In this section we will introduce cuts in our proof nets.
A \dfn{cut} is a formula $A\lcut A\lneg$, 
where  $\lcut$ is called the \dfn{cut connective}, and
where the function $\fneg$ is defined on formulas as follows (with an obvious
abuse of notation):
\begin{equation}
  \label{eq:neg}
\begin{array}{r@{\;=\;}l@{\qquad}r@{\;=\;}l@{\quad\qquad}r@{\;=\;}l}
a\lnegneg&a&\lone\lneg&\lbot&(A\ltens B)\lneg&B\lneg\lpar A\lneg\\
a\lneg&a\lneg&\lbot\lneg&\lone&(A\lpar B)\lneg&B\lneg\ltens A\lneg
\end{array}
\end{equation}
Notice that we invert the order under a negation, as if the logic
were not commutative. This considerably simplifies many proof nets and
categorical diagrams.

A \dfn{sequent with cuts} is a sequent where some of the formulas are
cuts. But cuts are not allowed to occur inside formulas, \ie all
$\lcut$-nodes are roots.  A \dfn{pre-proof graph with cuts} is a
pre-proof graph $\prfgrf{P}{\Gamma}$, where $\Gamma$ may contain cuts.
The $\lcut$-nodes have the same geometric behavior as the
$\ltens$-nodes. Therefore the correctness criterion stays literally
the same, and we can define \dfn{proof graphs with cuts} and
\dfn{proof nets with cuts} accordingly. In the translation from
sequent proofs containing the cut rule into pre-proof graphs with
cuts, the cut is treated as follows:
$$
\vciinf{\cutr}{\Gamma,\Delta}{\Gamma,A\;}{\;A\lneg,\Delta}
\quad\leadsto\quad
\vciinf{\cutr}{\prfgrf{P\lpar Q}{\Gamma,A\lcut A\lneg,\Delta}}
{\prfgrf{P}{\Gamma,A}\;}{\;\prfgrf{Q}{A\lneg,\Delta}}
\quad.
$$
Since the $\lcut$ behaves in the same way as the $\ltens$, we
immediately have the generalization of the sequentialization theorem:

\begin{theorem}\label{thm:cutsequent}
  A pre-proof net with cuts is sequentializable if and only if it is
  correct, \ie it is a proof net with cuts.
\end{theorem}

\proof
  This proof is literally the same as the proof of
  Theorem~\ref{thm:sequent}, with the only difference, that there are
  now also $\lcut$-nodes, which are treated as $\ltens$-nodes. \qed

\begin{remark}\label{rem:unitfreecut}
  In the same way, we can add the cut to ordinary proof nets, as
  defined in \ref{par:unitfree}. Of course, this does not affect the
  sequentialization.
\end{remark}

\subsection{Cut elimination}\label{sec:cutelim}

The famous cut elimination theorem says that for any proof containing cuts
there is a cut-free proof of the same conclusion. 
For MLL sequent calculus proofs this is a well-known fact. Since we
have sequentialization for cut-free proof nets, as well as for proof
nets with cuts (Theorems~\ref{thm:sequent} and~\ref{thm:cutsequent}),
we can immediately conclude a cut elimination result for proof nets. 

In this section we will present a procedure that will eliminate the
cuts directly on the proof nets. More precisely, we will present a
strongly normalizing cut reduction relation. This means that to every
proof net with cuts a unique cut free proof net is assigned.
 
On the set of cut pre-proof graphs we can define the cut 
reduction relation~$\prfred$ as follows:
$$
\begin{array}{r@{\;\;}c@{\;\;}l}
\prfgrfD{P}{(A\lpar B)\lcut(B\lneg\ltens A\lneg),\Gamma}&\prfred&
\prfgrfD{P}{A\lcut A\lneg,B\lcut B\lneg,\Gamma}\\ \\
\prfgrfD{P\cons{(a_h\lneg\ltens a_i)\lpar(a_j\lneg\ltens a_k)}}
{a_i\lcut a_j\lneg,\Gamma}&\prfred&
\prfgrfD{P\cons{a_h\lneg\ltens a_k}}{\Gamma}\\ \\
\prfgrfD{P\cons{(Q\ltens\lbot_i)\lpar\lone_j}}
{\lbot_i\lcut\lone_j,\Gamma}&\prfred&
\prfgrfD{P\cons{Q}}{\Gamma}
\end{array}
$$

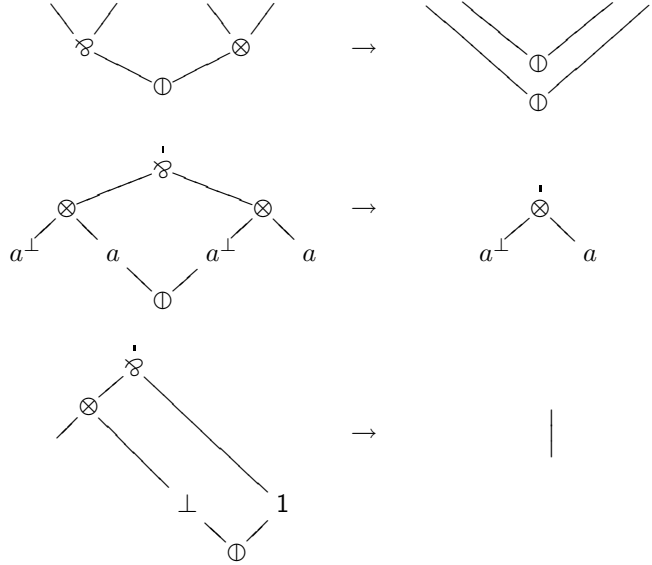
\begin{figure}[t]
\vskip-3ex
\small
  $$
  \begin{array}{ccc}
  \xymatrix@C=\matrcol @R=\matrrowa{\ar@{-}
    &&&&&&\\
    \\
    &\mpar{uur}{uul}&&&&\mtens{uur}{uul}\\
    &&&\mcut{urr}{ull}
  }
  &\lower2em\hbox{$\prfred$}&
  \xymatrix@C=\matrcol @R=\matrrowa{\ar@{-}
    \mnoat&&\mnoat&&\mnoat&&\mnoat\\
    \\
    &&&\mcut{uull}{uurr}\\
    &&&\mcut{uuulll}{uuurrr}\\
  }
  \\[-3ex]
  \vcenter{
  \xymatrix@C=\matrcol @R=\matrrowa{\ar@{-}
    \mnix&&\mnix&\mnix\mdown&\mnix&&\mnix\\
    &&&\mpar{dll}{drr}\\
    &\mtens{dl}{dr}&&&&\mtens{dl}{dr}\\
    \matn{a}\,\strut&&\mat{a}&&\matn{a}\,&&\mat{a}\\
    &&&\mcut{ul}{ur}\\
  }}
  &\prfred&
  \vcenter{
  \xymatrix@C=\matrcol @R=\matrrowa{\ar@{-}
    \mnix&\mnix&\mnix\\
    \mnix&\mnix\mdown&\mnix\\
    &\mtens{dl}{dr}\\
    \matn{a}\,&&\mat{a}\\
    \mnix&\\
  }}
  \\[-3ex]
  \vcenter{
  \xymatrix@C=\matrcol @R=\matrrowa{\ar@{-}
    &&\mnix\mdown&\\
    &&\mpar{dl}{dddrrr}\\
    &\mtens{dl}{ddrr}&\\
    \mnix&\\
    &&&\mbot&&\mone\\
    &&&&\mcut{ul}{ur}\\
  }}
  &\prfred&
  \quad\Big\vert
  \\[-2ex]
  \end{array}
  $$
   \caption{Cut elimination reduction steps}
   \label{fig:cutelim}
\end{figure}

These reduction steps are 
shown in graphical notation in Figure~\ref{fig:cutelim}.

We have the following immediate lemma, which ensures that correctness is
preserved during the reduction.

\begin{lemma}\label{lem:redcorrect}
  If $\prfgrf{P}{\Gamma}$ is correct and
  $\prfgrf{P}{\Gamma}\prfred\prfgrf{P'}{\Gamma'}$, then
  $\prfgrf{P'}{\Gamma'}$ is also correct.
\end{lemma}

\proof
It is impossible that a cut reduction step
introduce a cycle in a switching or make it disconnected.
\qed

Observe that it can happen that in a proof graph no
reduction is possible, although there are cuts present in the sequent.
For example, in 
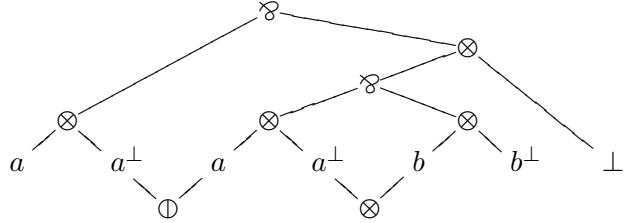
{
$$
  \xymatrix@C=\matrcol @R=\matrrow{\ar@{-}
    &&&&&\mpar{dddllll}{drrrr}\\
    &&&&&&&&&\mtens{dll}{dddrrr}\\
    &&&&&&&\mpar{dll}{drr}\\
    &\mtens{dl}{dr}&&&&\mtens{dl}{dr}&&&&\mtens{dl}{dr}\\
    \mat{a}&&\matn{a}&&\mat{a}&&\matn{a}&&\mat{b}&&\matn{b}&&\mbot\\
    &&&\mcut{ul}{ur}&&&&\mtens{ul}{ur}\\
  }
$$
}
the cut cannot be reduced. 

In a given proof graph $\prfgrf{P}{\Gamma}$, a $\lcut$-node that can
be reduced will be called \dfn{ready}. Obviously, a cut on a
$\ltens$-$\lpar$-pair is always ready, but for a cut on atoms or units
this is not necessarily the case, as the example above shows. 
However, we have the following theorem.

\begin{theorem}\label{thm:netred}
Given a proof graph $\prfgrf{P}{\Gamma}$ and a $\lcut$-node
in $\Gamma$, there is an equivalent
proof graph $\prfgrf{P'}{\Gamma}$, in which that $\lcut$-node is
ready, \ie can be reduced. 
\end{theorem}

This is an immediate consequence of the following two lemmas.

\begin{lemma}\label{lem:netredatom}
  For every proof graph $\prfgrf{P}{a_i\lcut a_j\lneg,\Gamma}$ that
  contains an atomic cut, there is an equivalent proof graph 
  $\prfgrfT{P'\cons{(a_h\lneg\ltens a_i)\lpar(a_j\lneg\ltens a_k)}}
  {a_i\lcut a_j\lneg,\Gamma}$.
\end{lemma}

\begin{lemma}\label{lem:netredunit}
  For every proof graph $\prfgrf{P}{\lbot_i\lcut\lone_j,\Gamma}$ that
  contains a cut on the units, there is an equivalent proof graph 
  $\prfgrfT{P'\cons{(Q\ltens\lbot_i)\lpar\lone_j}}
  {\lbot_i\lcut\lone_j,\Gamma}$.
\end{lemma}

For proving them, we will use the following three lemmas
(\ref{lem:deepa}--\ref{lem:deepc}).

\begin{lemma}\label{lem:deepa}
  Let 
  $\prfgrfT{P\cons{(\lbot_k\ltens R\cons{x_i})\lpar 
      (S\cons{x_j\lneg}\ltens\lbot_h)}}{x_i\lcut x_j\lneg,\Gamma}$ 
  be a proof graph, where $x$ is an arbitrary atom or a unit, and
  $x\lneg$ its dual.\\
  Then at least one of 
  $\prfgrfT{P\cons{\lbot_k\ltens(R\cons{x_i}\lpar 
      (S\cons{x_j\lneg}\ltens\lbot_h))}}{x_i\lcut x_j\lneg,\Gamma}$
  and\\
  $\prfgrfT{P\cons{((\lbot_k\ltens R\cons{x_i})\lpar 
      S\cons{x_j\lneg})\ltens\lbot_h}}{x_i\lcut x_j\lneg,\Gamma}$ 
  is equivalent to it.
\end{lemma}

\proof
\Bwoc, assume that both are not equivalent to the original proof
graph. This means that in both cases the side condition $(*)$
of~\ref{par:grfequ} is not fulfilled, which means that in the
original proof graph we have
\begin{itemize}
\item an extended switching wrt.~$\bot_k$ such that one node of
  $S\cons{x_j\lneg}\ltens\lbot_h$ is connected to it, and
\item an extended switching wrt.~$\bot_h$ such that one node of
  $\lbot_k\ltens R\cons{x_i}$ is connected to it.
\end{itemize}
\Wolg, we can assume that in both extended switching the $\bot_k$
(respectively $\bot_h$) is connected to the $\ltens$-root of
$S\cons{x_j\lneg}\ltens\lbot_h$ (respectively $\lbot_k\ltens
R\cons{x_i}$).  If the two paths do not have a common node, then we
can immediately construct a (normal, non-extended) switching in which
both are present. But this switching contains a cycle in which the two
paths are connected by the $\ltens$-roots of $\lbot_k\ltens
R\cons{x_i}$ and $S\cons{x_j\lneg}\ltens\lbot_h$, contradicting the
assumption of correctness.  If the two paths have at least one common
node, we can also construct a switching with a cycle as follows. We
can make sure that it contains the first path between $\bot_k$ and the
first intersection node with the second path. Note that is path must
go ``downwards'' from the $\bot_k$ because the edge between the
$\bot_k$ and the root of $\lbot_k\ltens R\cons{x_i}$ is not present in
the extended switching.  Then the next node of the first path does
also belong to the second path (otherwise we would have a graph node
with four edges attached to it). We can now make sure that that part
of the second path, which is determined by the direction established
by these two nodes, is contained in the switching. There are now two
possibilities:
\begin{itemize}
\item We get a path between $\lbot_k$ and the $\ltens$-root of
  $\lbot_k\ltens R\cons{x_i}$, which yields a cycle
  immediately because now the edge between the $\bot_k$ and its
  $\ltens$-parent is present.
\item We get a path between $\lbot_k$ and $\bot_h$. There are two
  subcases:
  \begin{itemize}
  \item The path does not contain nodes from $S\cons{x_j\lneg}$. 
    In this case we can extend the path to a cycle using the two
    $\ltens$-roots of $\lbot_k\ltens R\cons{x_i}$ and
    $S\cons{x_j\lneg}\ltens\lbot_h$, as well as 
    the $\lcut$-node between $x_i$ and $x_j\lneg$.
  \item The path does contain nodes from $S\cons{x_j\lneg}$. We
    consider the last part of this path which connects a leaf of
    $S\cons{x_j\lneg}$ with $\bot_h$, without touching any other node of
    $S\cons{x_j\lneg}$, and which does not contain the $\ltens$-root of
    $S\cons{x_j\lneg}\ltens\lbot_h$ because the edge between $\bot_h$
    and its $\ltens$-parent is not present in the extended switching.
    This path can now be extended to a cycle that contains
    $\ltens$-parent of $\bot_h$.
  \end{itemize}
\end{itemize}
This contradicts the assumption of correctness of the original graph.
\qed

\begin{lemma}\label{lem:deepb}
  Let 
  $\prfgrfT{P\cons{(\lbot_k\ltens R\cons{x_i})\lpar (x_j\lneg\ltens Q)}}
  {x_i\lcut x_j\lneg,\Gamma}$ 
  be a proof graph, where $x$ is an arbitrary atom or a unit, and
  $x\lneg$ its dual.\\
  Then $\prfgrfT{P\cons{\lbot_k\ltens(R\cons{x_i}\lpar (x_j\lneg\ltens Q))}}
  {x_i\lcut x_j\lneg,\Gamma}$ 
  is equivalent to it.
\end{lemma}

\proof
  \Bwoc, assume this is not the case, \ie the side condition $(*)$
  of~\ref{par:grfequ} is not fulfilled, which means that we have in
  the original proof graph an extended switching wrt.~$\bot_k$ such
  that a node of $x_j\lneg\ltens Q$ is connected to it.  If this path
  goes through $R\cons{x_i}$, we have a cycle immediately.  If this is
  not the case, it has to enter $x_j\lneg\ltens Q$ either from the
  $\lpar$-node above or through a leaf of $Q$. In both cases we can
  extend the path through $x_j\lneg$ and the $\lcut$-node to $x_i$ and
  the root of $R\cons{x_i}$, which yields a cycle.  \qed

\begin{lemma}\label{lem:deepc}
  Let 
  $\prfgrfT{P\cons{(\lbot_k\ltens R\cons{x_i})\lpar x_j\lneg}}
  {x_i\lcut x_j\lneg,\Gamma}$ 
  be a proof graph, where $x$ is an arbitrary atom or a unit, and
  $x\lneg$ its dual.\footnote{In fact, according to the definition of proof
  graph, the only possibility here is that $x=\lbot$ and
  $x\lneg=\lone$. However, in order to emphasize a certain uniformity in all
  three lemmas, we use $x$.}\\
  Then $\prfgrfT{P\cons{\lbot_k\ltens(R\cons{x_i}\lpar x_j\lneg)}}
  {x_i\lcut x_j\lneg,\Gamma}$ 
  is equivalent to it.
\end{lemma}

\proof
Similar to Lemma~\ref{lem:deepb}.
\qed

We can now proceed with the proofs of Lemmas \ref{lem:netredatom} 
and~\ref{lem:netredunit}.

\proof[Proof of Lemma~\ref{lem:netredatom}]
By the definition of proof graph, the linking $P$ must contain two subtrees
$a_h\lneg\ltens a_i$ and $a_j\lneg\ltens a_k$. By the correctness
criterion, they must be in a $\lpar$-relation, \ie
$P=P''\cons{R\cons{a_h\lneg\ltens a_i}\lpar S\cons{a_j\lneg\ltens a_k}}$ 
for some contexts $P''\conhole$, $R\conhole$ and $S\conhole$.
We will proceed by induction on the size of $R\conhole$ and
$S\conhole$, \ie the sum of the
number of $\lpar$- and $\ltens$-nodes in them. We have the following cases:
\begin{itemize}
\item Both are empty. In this case we are done.
\item $R\conhole$ has a $\ltens$ as root and $S\conhole$ is empty. In
  this case $R\conhole=\lbot\ltens R'\conhole$ for some $R'\conhole$,
  and we can apply Lemma~\ref{lem:deepb} with $Q=a_k$.
\item $R\conhole$ is empty and $S\conhole$ has a $\ltens$ as
  root. This case is symmetrical to the previous one, 
  and we can apply Lemma~\ref{lem:deepb} with $Q=a_h\lneg$.
\item Both $R\conhole$ and $S\conhole$ have a $\ltens$ as root. In
  this case, we can apply Lemma~\ref{lem:deepa}, and proceed by
  induction hypothesis.
\item One of $R\conhole$ and $S\conhole$ has a $\lpar$ as root. In
  this case we apply the associativity of the $\lpar$ (which is not
  subject to a side condition), and proceed by
  induction hypothesis.  \qed
\end{itemize}

\proof[Proof of Lemma~\ref{lem:netredunit}]
  This proof is very similar to the previous one. Since
  $\prfgrf{P}{\lbot_i\lcut\lone_j,\Gamma}$ is correct,
  it is of the shape
  $\prfgrfT{P''\cons{R\cons{Q\ltens\lbot_i}\lpar
      S\cons{\lone_j}}}{\lbot_i\lcut\lone_j,\Gamma}$.  
  Again, we will proceed by induction
  on the size of $R\conhole$ and $S\conhole$, with almost identical
  cases.  The only difference is:
\begin{itemize}
\item $R\conhole$ has a $\ltens$ as root and $S\conhole$ is empty. In
  this case we apply Lemma~\ref{lem:deepc} (instead of
  Lemma~\ref{lem:deepb}).
  \qed
\end{itemize}

This completes the proof of Theorem~\ref{thm:netred}.

Let us now extend the relation $\prfred$ to proof nets as follows: We say
$$\prfnet{P}{\Gamma}\prfred\prfnet{Q}{\Delta}$$ 
if an only if there are
proof graphs $\prfgrf{P'}{\Gamma}$ and $\prfgrf{Q'}{\Delta}$ such
that
$$
\prfgrfT{P}{\Gamma}\prfequ\prfgrfT{P'}{\Gamma}\prfred
\prfgrfT{Q'}{\Delta}\prfequ\prfgrfT{Q}{\Delta}
\quadfs
$$

Let us first show that this is well-defined, in the sense that if the
same cut is reduced in two different representatives of the same net,
then the two results do also represent the same net.

\begin{lemma}\label{lem:redinvar}
  Let $\prfgrf{P}{\Gamma}\prfequ\prfgrf{P'}{\Gamma}$, and let 
  $\prfgrfT{P}{\Gamma}\prfred\prfgrfT{Q}{\Delta}$ and 
  $\prfgrfT{P'}{\Gamma}\prfred\prfgrfT{Q'}{\Delta}$,
  \ie in both
  reductions the same cut is reduced. Then we have 
  $\prfgrf{Q}{\Delta}\prfequ\prfgrf{Q'}{\Delta}$.
\end{lemma}

\proof
Since $\prfgrf{P}{\Gamma}\prfequ\prfgrf{P'}{\Gamma}$, we have 
$$
\prfgrfT{P}{\Gamma}=\prfgrfT{P_0}{\Gamma}\prfequ\prfgrfT{P_1}{\Gamma}
\prfequ\cdots\prfequ\prfgrfT{P_n}{\Gamma}=\prfgrfT{P'}{\Gamma}
\;,
$$
for some linkings $P_0$, $P_1$, \ldots, $P_n$, where for each
$i=1,\ldots,n$ the equivalence
$\prfgrf{P_{i-1}}{\Gamma}\prfequ\prfgrf{P_i}{\Gamma}$ is a direct
application of the equations in~\ref{par:grfequ}.  We can now
distinguish three cases.

First, the reduced cut is on
binary connectives. Then in each of the proof graphs
$\prfgrf{P_i}{\Gamma}$ the cut is ready and we have
$
\prfgrfT{Q}{\Delta}=\prfgrfT{Q_0}{\Delta}
\prfequ\cdots\prfequ\prfgrfT{Q_n}{\Delta}=\prfgrfT{Q'}{\Delta}
,
$
where each $\prfgrfT{Q_i}{\Delta}$ is obtained from reducing the cut
in $\prfgrf{P_i}{\Gamma}$.

In the second case the reduced cut is an atomic one, say $a_i\lcut
a_j\lneg$. Here it might happen that in some of the
$\prfgrf{P_i}{\Gamma}$ the cut is not ready because of unnecessary
applications of associativity. But it is easy to see that there is a
transformation from $\prfgrf{P}{\Gamma}$ to $\prfgrf{P'}{\Gamma}$ in
which the readiness of the cut is not destroyed, \ie the sublinking
$(a_h\lneg\ltens a_i)\lpar(a_j\lneg\ltens a_k)$ of $P$ and $P'$ is not
touched. We can therefore proceed as in the first case.

The most difficult case occurs if the cut is on the units, say
$\lbot_i\lcut\lone_j$. Although
$P=R\cons{(S\ltens\lbot_i)\lpar\lone_j}$ and
$P'=R'\cons{(S'\ltens\lbot_i)\lpar\lone_j}$, the sublinking
$(-\ltens\lbot_i)\lpar\lone_j$ might be destroyed in the
transformation because other subtrees might leave or enter the scope
of the $\lbot_i$, and can therefore occur ``between'' $\lbot_i$ and
$\lone_j$. However, in the reduction $\lbot_i$ and $\lone_j$
disappear. Hence these intermediate steps become vacuous. We can
therefore proceed similarly to the other two cases.  \qed

The next thing to check is termination:

\begin{lemma}\label{lem:termin}
  There is no infinite sequence 
  $$
  \prfnet{P}{\Gamma}\prfred\prfnet{P'}{\Gamma'}\prfred
  \prfnet{P''}{\Gamma''}\prfred\cdots
  $$
\end{lemma}

\proof
In each reduction step the size of the sequent (\ie the number of
$\lpar$, $\ltens$ and $\lcut$-nodes) is reduced.
\qed

For showing confluence of the reduction relation, we will proceed in two
steps. First we will show that the reduction relation on (pre-)proof graphs is
confluent, and then we will extend the result to (pre-)proof nets, by
employing Lemma~\ref{lem:redinvar}.
    
\begin{lemma}\label{lem:grfconf}
  If $\prfgrf{Q}{\Delta}\redprf\prfgrf{P}{\Gamma}\prfred\prfgrf{R}{\Sigma}$,
  then either
  $\prfgrf{Q}{\Delta}=\prfgrf{R}{\Sigma}$, or there is a proof graph
  $\prfgrf{S}{\Phi}$ such that 
  $\prfgrf{Q}{\Delta}\prfred\prfgrf{S}{\Phi}\redprf\prfgrf{R}{\Sigma}$.
\end{lemma}

\proof
  If $\prfgrf{Q}{\Delta}$ and $\prfgrf{R}{\Sigma}$ are obtained from
  $\prfgrf{P}{\Gamma}$ by reducing the same $\lcut$-node in $\Gamma$
  then they must be equal. If different $\lcut$-nodes have been
  reduced in the two reductions, then both $\lcut$-nodes must have
  been ready in $\prfgrf{P}{\Gamma}$. But reducing one of the
  two $\lcut$-nodes does not destroy the readiness of the other, which
  can therefore be reduced afterwards. Since the redexes of the
  reductions cannot ``overlap'', the result $\prfgrf{S}{\Phi}$ 
  is independent of the order of
  the two reductions. 
  \qed

\begin{lemma}\label{lem:netconf}
  If
  $\prfnet{Q}{\Delta}\redprf\prfnet{P}{\Gamma}\prfred\prfnet{R}{\Sigma}$,
  then either $\prfnet{Q}{\Delta}=\prfnet{R}{\Sigma}$, or there is a
  proof net $\prfnet{S}{\Phi}$ such that
  $\prfnet{Q}{\Delta}\prfred\prfnet{S}{\Phi}\redprf\prfnet{R}{\Sigma}$.
\end{lemma}

\proof
  The problem is that the two reduction might take place in two
  different presentations of the proof net $\prfnet{P}{\Gamma}$.
  (Otherwise we could immediately apply Lemma~\ref{lem:grfconf}.) The
  main idea of this proof is therefore to exhibit a presentation of
  $\prfnet{P}{\Gamma}$ in which both cuts are ready, and then apply
  Lemma~\ref{lem:grfconf} together with Lemma~\ref{lem:redinvar}.  Let
  $\lcut_1$ denote the cut that is reduced in $\Gamma$ to obtain
  $\Delta$ and $\lcut_2$ the one that is reduced to obtain $\Sigma$.
  If they are identical, we immediately have (by
  Lemma~\ref{lem:redinvar}) that
  $\prfnet{Q}{\Delta}=\prfnet{R}{\Sigma}$. If not, we distinguish the
  following cases:
  \begin{itemize}
    \item One of the two cuts is on binary connectives, \ie it is
      ready in each presentation of the proof net. We can therefore
      choose a presentation in which the other cut is also ready and
      apply Lemma~\ref{lem:grfconf} and Lemma~\ref{lem:redinvar}.
    \item One of the two cuts is on units, say $\lcut_1$. Then we can
      first make $\lcut_2$ ready by applying Lemma~\ref{lem:netredatom}
      or Lemma~\ref{lem:netredunit}. Then we apply
      Lemma~\ref{lem:netredunit} to also make $\lcut_1$ ready. This
      does not affect the readiness of $\lcut_2$. We can
      therefore obtain a presentation of $\prfnet{P}{\Gamma}$ in which
      both cuts are ready, and proceed as in the previous case.
    \item Both cuts are atomic, but are not
      directly connected to each other via a ``real'' axiom link.
      Then we can
      proceed as in the previous case to obtain a  
      presentation of $\prfnet{P}{\Gamma}$ in which both cuts are
      ready.
    \item Both cuts are atomic and share a common ``real'' axiom link. In other
      words, $\prfgrf{P}{\Gamma}$ is of the following shape:
      $$
      {\prfgrfD{P'\cons{(P''\cons{a_h\lneg\ltens a_i}\lpar
          P'''\cons{a_j\lneg\ltens a_k})\lpar 
          P''''\cons{a_l\lneg\ltens a_m}}}
             {a_i\lcut_1 a_j\lneg,a_k\lcut_2 a_l\lneg,\Phi}}
      $$
      In this case it is not possible to make both cuts ready at the
      same time. But we can transform the above graph into
      $$
      \prfgrf{S'\cons{((a_h\lneg\ltens a_i)\lpar
          (a_j\lneg\ltens a_k))\lpar 
          (a_l\lneg\ltens a_m)}}
             {a_i\lcut_1 a_j\lneg,a_k\lcut_2 a_l\lneg,\Phi\quadcm}
      $$
        as well as into
      $$
      \prfgrf{S'\cons{(a_h\lneg\ltens a_i)\lpar
          ((a_j\lneg\ltens a_k)\lpar 
          (a_l\lneg\ltens a_m))}}
             {a_i\lcut_1 a_j\lneg,a_k\lcut_2 a_l\lneg,\Phi\quadfs}
      $$
      In the first case $\lcut_1$ is ready and in the second
      $\lcut_2$. In both cases, after the reduction of one cut,
      the other becomes ready. After
      the second reduction, the result is in both cases
      $\prfgrf{S'\cons{a_h\lneg\ltens a_m}}{\Phi}$.
      \qed
  \end{itemize}

\begin{theorem}\label{thm:strnorm}
The cut elimination 
reduction $\prfred$ on proof nets is strongly normalizing. The
normal forms are cut free proof nets.
\end{theorem}

\proof
   Termination is provided by Lemma~\ref{lem:termin}, confluence
   follows from Lemma~\ref{lem:netconf}, and that the normal form is cut
   free is ensured by Theorem~\ref{thm:netred}.
   \qed

\begin{figure}[t]
\vskip-3ex
\small
  $$
  \begin{array}{ccc}
  \xymatrix@C=\matrcol @R=\matrrowa{\ar@{-}
    &&&&&&\\
    \\
    &\mpar{uur}{uul}&&&&\mtens{uur}{uul}\\
    &&&\mcut{urr}{ull}
  }
  &\lower2em\hbox{$\prfred$}&
  \xymatrix@C=\matrcol @R=\matrrowa{\ar@{-}
    \mnoat&&\mnoat&&\mnoat&&\mnoat\\
    \\
    &&&\mcut{uull}{uurr}\\
    &&&\mcut{uuulll}{uuurrr}\\
  }
  \\
\\
  \vcenter{
  \begin{xy}
    0;<2em,0ex>:<0em,4ex>::
    (0,1)="a"*{\sa}, (2,1)="b"*{\sna}, (4,1)="c"*{\sa},
    (6,1)="d"*{\sna}, 
    (3,0)="h"*{\lcut}, 
    "b"*{\ptens};"h"*{\ptens} **\dir{-},
    "c"*{\ptens};"h"*{\ptens} **\dir{-},
    "a"*{\ptens};"b"*{\ptens} **\crv{~*=<.5pt>{.}(1,\doubleheight)},
    "c"*{\ptens};"d"*{\ptens} **\crv{~*=<.5pt>{.}(5,\doubleheight)},
  \end{xy}
  }
  &\prfred&
  \vcenter{
  \begin{xy}
    0;<2em,0ex>:<0em,4ex>::
    (1,1)="a"*{\sa}, 
    (5,1)="d"*{\sna}, 
    (3,0)="h"*{\ptens}, 
    "a"*{\ptens};"d"*{\ptens} **\crv{~*=<.5pt>{.}(3,\doubleheight)},
  \end{xy}
  }
  \\
  \end{array}
  $$
   \caption{Cut elimination reduction steps for ordinary proof nets}
   \label{fig:unitfreecutelim}
\end{figure}
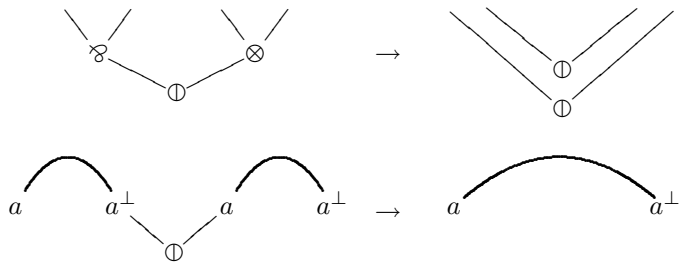

\begin{para}{\bf Cut elimination for ordinary proof nets.}%
  \label{par:unitfreecutelim}
  The following is very well known (see \eg
  \cite{girard:87,danos:regnier:89,retore:03}), but we add it for the
  sake of completeness.  Define the cut reduction relation on the set
  of ordinary pre-proof nets as shown in
  Figure~\ref{fig:unitfreecutelim}.  There are only two cases: the cut
  on binary connectives and the cut on atoms. A cut on binary
  connectives is replaced by two cuts on the corresponding subformulas
  (as in the case with units), and a cut on atoms is removed by
  melting the two attached axiom links to a single axiom link. It is
  easy to see that this reduction preserves correctness, and is
  terminating and confluent. Therefore Theorem~\ref{thm:strnorm} does
  also hold for ordinary proof nets.
\end{para}

\section{*-Autonomous categories}\label{sec:categories}

In the great majority of cases, a category with additional structure
turns out to be a category that obeys a certain class of universal
properties: products, coproducts, right adjoint to products,
equalizers, etc. That is, the structure in question ends up being a
\emph{property} of the category, from which operations, in a more
standard algebraic sense, can be extracted via the axiom of choice.
There are exceptions to this, and the most important one, by far, is
the concept of monoidal structure, which cannot be defined abstractly
without recourse to an explicit binary operation, i.e., a bifunctor.

On this operation a form of associativity holds, which is not an
equation but a natural isomorphism. A unit object for it is
almost always present (hence the adjective {\em monoidal\/}), and very
often there is additional structure like a braiding or a symmetry,
that corresponds to a suitably generalized form of commutativity.

Monoidal categories abound in nature, and the first examples were in
the world of rings and modules (and in the closely related world of
topological vector spaces): the category of modules over a commutative
ring---thus Abelian groups form an important special case of
this---has a symmetric monoidal structure given by the operation of
tensoring.  The category of left-right bimodules over an arbitrary
ring has a monoidal structure, not symmetrical in general. But
bimodules have another interesting binary operation, the bimodule of
functions, which obeys a relation of adjointness with the tensor. This
additional structure on modules led to the axiomatization of monoidal
closed categories. Even before these abstract concepts of monoidal
and monoidal closed categories had been formulated,
Lambek~\cite{lambek:calcsynt} had noticed the strong logical flavor of
these operations, the function module operation being a form of
implication, whose associated conjunction, tensoring, was not
necessarily commutative.

Monoidal closed categories first appeared in
\cite{eilenberg:kelly:closed-cat}, and they give the necessary axiomatic
treatment for Lambek's implication; as a matter of fact they were
formulated in such a way that in some cases an implication/function
object operator can be defined without the presence of a
conjunction/tensor. Thus there can be closed categories that are
not monoidal.

Constructing free categories-with-structure is an interesting problem
by itself; for historical reasons it is called ``solving a coherence
problem''. So as soon as the concept of closed category was
formulated, there was the question of describing free ones. Lambek
immediately saw the relationship between monoidal closed categories
and logic, and produced a specific cut-elimination theorem soon after
they were introduced~\cite{lambek:68,lambek:69}. Thus, from the start
the relationship between logic and categories was bidirectional.
Abstract properties of semantical categories could give a way of
formulating semantics for logical systems, as well as suggesting new
such systems. And logical tools like cut-elimination could help the
construction of free categories-with-structure. As a matter of fact
the problem of getting a complete description of the free monoidal
closed category, by means logical or not, has led to a sizable lot of
publications over the last 35 years.

But some categories of modules have even more structure. First, for
any commutative ring \(R\) there is always a notion of dual: the dual
\(M^{\ast}\) of an \(R\)-module \(M\) is given by taking the module of
functions of \(M\) into \(R\). This defines a contravariant
endofunctor; but, moreover, if \(R\) is a field and if we restrict
ourselves to finitely generated modules (= finite dimensional vector
spaces) over that field, then we get that \(M\) and the bidual
\(M^{**}\) are related by a natural isomorphism. Thus we have
something very much like an involutive negation in logic\ldots\ except
that in this particular case the ``false'' object \(R\) also takes the
role of ``true'' in Lambek's logical interpretation.

Driven by purely algebraic considerations M.~Barr~\cite{barr:79}
started looking for more examples of symmetric monoidal closed
categories that have such a ``dualizing'' object, i.e., where every
object is naturally isomorphic to its bidual. This led him to the
formulation of *-autonomous categories, and to examples where the
dualizer was not necessarily the unit to the tensor. He also found a
general technique for producing such categories out of any ordinary
monoidal closed category that has pullbacks; this is now called the
Chu construction~\cite{chu:79}. The main inspiration for the Chu
construction had been around since Mackey's thesis of 1942, published
in 1945~\cite{mackey:45}, and can be summarized as follows. In the
realm of topological vector spaces, it is very desirable to have a
notion of duality which is naturally involutive, as above; the
canonical example is the category of Hilbert spaces, which is
unfortunately a very restricted case.  Mackey's idea was to decide
that a topology on a vector space over a complete normed field could
be replaced by an abstract notion of dual: another vector space whose
elements are to be seen as continuous linear functionals for the other
space. The relationship between these two spaces is quite symmetrical,
and so the operation of taking the dual simply becomes the exchange of
the two spaces.

The discovery of linear logic by Girard was completely independent
from this, but came from the observation of a particular *-autonomous
category, that of coherence spaces and linear maps. Coherence spaces
are more closely related to the category of sets and relations than to
the category of Abelian groups, but from the beginning Girard was
aware that there were numerous points of contact between linear
algebra and the improved logic he was seeking to create.
Hence his choice of the name ``linear logic''. 
It did not take long to establish that the categorical framework for
axiomatizing (multiplicative) linear logic was *-autonomous
categories; that was worked out in~\cite{lafont:thesis,seely:89}.
Particular cases of the Chu construction were then rediscovered by the
linear logicians: applying Chu to sets and functions gives the
category of Lafont-Streicher games~\cite{lafont:streicher:91} (now
better known as ``Chu spaces''), and applying it to Banach spaces
gives Girard's ``coherent Banach spaces''~\cite{girard:96:Banach}, thus
closing a fifty-year loop.

\subsection{Basic definitions and properties}\label{sec:defstar}

In this section we will recall the definition of a *-autonomous
category and show some properties that they have and that we will use.
We assume that the reader is familiar with the basic notions of
category theory. Given a category \(\CC\) and maps \(f\colon A\to B\)
and \(g\colon B\to C\) in \(\CC\) we will write the composition of
\(f, g\) either as \(gf\) or \(g\circ f\), depending on the needs of
readability. We will write the identity map on, say \(A\) as
\(\id_{A}\), although there is much to be said for writing it as
simply \(A\). We do not do this here because it tends to confuse
beginners.

\begin{definition}\label{def:monoidal}
  A \dfn{monoidal category} is a category $\CC$, equipped with a
  bifunctor $-\ltens-:\CC\times\CC\to\CC$, a distinguished object
  $\lone_\CC$, called the 
  \dfn{unit object}\footnote{The choice of typeface should prevent confusion
    with identity maps.}, and for all
  objects $A$, $B$, $C$ natural isomorphisms
  \begin{eqnarray*}
    \alpha_{A,B,C}&:& A\ltens(B\ltens C)\to(A\ltens B)\ltens C\quadcm\\
    \rho_A&:& A\ltens\lone_\CC\to A\quadcm\\
    \lambda_A&:& \lone_\CC\ltens A\to A\quadcm
  \end{eqnarray*}
  such that the following diagrams commute:
  $$
  \xymatrix@C=\diahcol@R=\diarow{
    A\ltens(\lone_\CC\ltens B)
    \ar[rr]^{\alpha_{A,\lone_\CC,B}}
    \ar[dr]_{\id_A\ltens \lambda_B}&&
    (A\ltens\lone_\CC)\ltens B
    \ar[dl]^{\rho_A\ltens \id_B}\\
    &\qqlapm{A\ltens B}
  }
  $$
  $$
  \xymatrix@C=\diazcol@R=\diarow{
    &&\qqqqlapm{(A\ltens B)\ltens (C\ltens D)}
    \ar[drr]^{\alpha_{A\ltens B,C,D}}\\
    \qqqlapm{A\ltens(B\ltens (C\ltens D))}^{\strut}
    \ar[urr]^{\alpha_{A,B,C\ltens D}}
    \ar[dr]_{\id_A\ltens \alpha_{B,C,D}}&&&&
    \qqqlapm{((A\ltens B)\ltens C)\ltens D}^{\strut}\\
    &A\ltens ((B\ltens C)\ltens D)\ar[rr]_{\alpha_{A,B\ltens C,D}}&
    &(A\ltens (B\ltens C))\ltens D\ar[ur]_{\alpha_{A,B,C}\ltens \id_D}
  }
  $$
\end{definition}
Let us introduce notation that will be very useful.  Let $I$ be a
finite index set. A \dfn{bracketing} of $I$ is given by a total order
on $I=\set{i_1,\ldots,i_k}$ and a binary tree with $k$ leaves indexed
by $I$ such that the order is respected.  We will denote bracketings
of $I$ also by $I$. Thus, given an $I$-indexed family $(C_{i})_{i\in
  I}$ of objects of $\CC$, we can use the notation $
\lbigtens{I}{\set{C_{i_1},\ldots,C_{i_k}}} $ to denote the object of
$\CC$ that is obtained by applying the functor $-\ltens-$ according to
the bracketing $I$. For empty $I$, let
$\lbigtens{\emptyset}{\emptyset}=\loneC$.  This allows us to state the
following proposition, which we will need later.

\begin{proposition}\label{prop:moncoh}
  Let $I$ be a finite index set, and let $I'$ and $I''$ be two
  bracketings on~$I$ that share the same order on
  $I=\set{i_1,\ldots,i_k}$. 
  Then for every
  $I$-indexed family $(C_{i})_{i\in I}$ of objects of a monoidal
  category $\CC$, there is a uniquely determined natural isomorphism
  $$\phi:\lbigtens{I'}{\bigset{C_i\mid i\in I}}\to
  \lbigtens{I''}{\bigset{C_i\mid i\in I}}\quadcm$$
  constructed only with the available data.
\end{proposition}

\proof
  This is an immediate consequence of the well-known coherence theorem
  for monoidal categories. (See, \eg \cite[Chapter VII.2]{maclane:71})
  \qed
As a matter of fact, we only have used \emph{part} of the coherence
theorem, the part that deals only with the tensor. What the above says
is that we can drop parentheses when we write an expression involving
only the tensor operation and an arbitrary family of objects. We just
have to make sure the objects are in the same order in both
expressions. But the full monoidal coherence theorem says more: not
only can we drop parentheses, and write a tensor of arbitrary objects
as a list/sequence, we also have the right to insert units anywhere we
want in that list. There will always be a \emph{unique} way to go from
one to the other via a coherent isomorphism.

\begin{definition}\label{def:symmetric}
  A \dfn{symmetric monoidal category} is a monoidal category \(\CC\)
  in which for all objects $A$ and $B$, there is a natural isomorphism
  (the \emph{symmetry})
  \begin{eqnarray*}
    \sigma_{A,B}&:& A\ltens B\to B\ltens A\quadcm
  \end{eqnarray*}
  such that the following diagrams commute:
  $$
  \xymatrix@C=\diahcol@R=\diarow{
    A\ltens B
    \ar[rr]^{\id_{A\ltens B}}
    \ar[dr]_{\sigma_{A,B}}&&
    A\ltens B\\
    &\qqlapm{B\ltens A}\ar[ur]_{\sigma_{B,A}}
  }
  $$
  $$
  \xymatrix@C=\diahcol@R=\diarow{
    A\ltens\lone_\CC
    \ar[rr]^{\sigma_{A,\lone_\CC}}
    \ar[dr]_{\rho_A}&&
    \lone_\CC\ltens A
    \ar[dl]^{\lambda_A}\\
    &A
  }
  $$
  $$
  \xymatrix@C=\diahcol@R=\diarow{
    &(A\ltens B)\ltens C\ar[rr]^{\sigma_{A\ltens B,C}}&
    &C\ltens (A\ltens B)\ar[dr]^{\alpha_{C,A,B}}\\
    A\ltens(B\ltens C)
    \ar[ur]^{\alpha_{A,B,C}}
    \ar[dr]_{\id_A\ltens \sigma_{B,C}}&&&&
    (C\ltens A)\ltens B\\
    &A\ltens (C\ltens B)\ar[rr]_{\alpha_{A,C,B}}&
    &(A\ltens C)\ltens B\ar[ur]_{\sigma_{A,C}\ltens \id_B}
  }
  $$
\end{definition}

Proposition~\ref{prop:moncoh} can be generalized to symmetric monoidal
categories, where we now drop the additional conditions that the two
bracketings $I'$ and $I''$ on $I$ have to share the same order.

\begin{proposition}\label{prop:symcoh}
  Let $I$ be a finite index set, and let $I'$ and $I''$ be two
  bracketings on~$I$. 
  Then for every
  $I$-indexed family $(C_{i})_{i\in I}$ of objects of a symmetric
  monoidal
  category~$\CC$, there is a uniquely determined natural isomorphism
  $$\phi:\lbigtens{I'}{\bigset{C_i\mid i\in I}}\to
  \lbigtens{I''}{\bigset{C_i\mid i\in I}}\quadfs$$
\end{proposition}

\proof
  As in the nonsymmetric case, the proposition is an immediate
  consequence of the coherence theorem, which for symmetric monoidal
  categories has first been proved in \cite{maclane:63}.  \qed
Thus, not only can we drop the parentheses in a tensor-unit
expression, now we can also change the order in which things are
written. But we have to be a bit more careful in the symmetrical case.
Given two expressions that involve the same family of objects, when
one of these objects appears more than once in the family, we have to
explicitly state how its instances are permuted between the two
expressions. An expression like \(A\ltens A \ltens A\) is used in
practice, but it should be more like \(A_{2}\ltens A_{1} \ltens
A_{3}\) or \(A_{3} \ltens A_{1} \ltens A_{2}\) or whatever.

\begin{definition}\label{def:staraut}
  \sloppy
  A \dfn{*-autonomous category} is a 
  symmetric monoidal category $\CC$ equipped with a
  a contravariant functor $\fneg:\CC\to\CC$, such that for any
  object $A$, we have a natural isomorphism $A\lnegneg\isom A$, 
  and for any three objects $A$,
  $B$, $C$ there is a natural bijection between
  \begin{eqnarray*}
    \Hom_\CC(A\ltens B,C)&\quad\mbox{and}\quad&\Hom_\CC(A,C\lpar B\lneg)
    \quadcm
  \end{eqnarray*}
  where the bifunctor $-\lpar-:\CC\times\CC\to\CC$ is defined by
  $A\lpar B=(B\lneg\ltens A\lneg)\lneg$.
  \footnote{Observe that we stick to our notational convention of 
    making negation switch the arguments; this is not strictly
    necessary but makes many formulas simpler to write.}
  We write $\lbot_\CC$ for $\lone_\CC\lneg$, and call it the
  \dfn{dualizing object} of \(\CC\).  We say a
  *-autonomous category is \dfn{strict} if the isomorphism
  $A\lnegneg\to A$ is an identity for all objects $A$.\footnote{Note
    that the symmetric monoidal structure is not necessarily strict
    (in the usual sense that associativity and symmetry are
    identities).}
\end{definition}

From now on \(\CC\) denotes a *-autonomous category.  A first
immediate consequence of this definition is that on every *-autonomous
category $\CC$ we have a second symmetric monoidal structure imposed
by the bifunctor $-\lpar-:\CC\times\CC\to\CC$ and its unit object
$\lbot_\CC$.  We will use the same notation (i.e.,
\(\alpha,\lambda,\rho,\sigma\)\ldots) for the natural isos associated
to that new symmetric monoidal structure, and we will also use
$\lbigpar{I}{\set{C_i\mid i\in I}}$ (where $I$ is a bracketing on an
index set and $(C_{i})_{i\in I}$ an $I$-indexed family) in the same
way as it has been done before for the bifunctor $-\ltens-$.  For
empty $I$, let $\lbigpar{\emptyset}{\emptyset}=\lbotC=\lone_\CC\lneg$.
Since there is no risk of confusion we tend to denote the objects
$\lone_{\CC}$ and~$\lbot_{\CC}$ by $\lone$ and~$\lbot$, respectively. 

Let us extract some standard consequences of that natural bijection
\begin{eqnarray}
  \Hom_\CC(A\ltens B\lneg,C)&\isom&\Hom_\CC(A,C\lpar B)
  \label{eq:biject}
  \quadfs
\end{eqnarray}
Notice that we have swapped \(B\) and \(B\lneg\), which is perfectly
legal given the involutory property of \((-)\lneg\); this way of
writing things is often more convenient for us. In the above, replace
\(A\) by \(C \lpar B\). In the right half plug the identity
\(\id_{C\lpar B}\), so at the left we get \(\epsilon_{B,C} \colon
(C \lpar B) \ltens B\lneg \to C\), the \dfn{evaluation map} which
obeys the well-known universal property:

\begin{proposition}\label{prop:basiccurry}
  Let $A$, $B$, $C$ be any objects of \(\CC\), and let \(f\in
  \Hom_{\CC} (A\ltens B\lneg, C)\) and \(g\in \Hom_{\CC}(A, C\lpar B)
  \) be related by the bijection\/ {\rm (\ref{eq:biject})}. Then \(g\)
  is the unique map such that \(f = \epsilon_{B,C} \circ(g\ltens
  \id_{B\lneg})\):
  \[ 
  \xymatrix@C=4em@R=5ex{ A \ltens B\lneg \ar[r]^-{g\ltens
      \id_{B\lneg}} & (C\lpar B)\ltens B\lneg \ar[r]^-{\epsilon_{B,C}}
    & C }
  \]
\end{proposition}

The proof can found in any textbook on category theory applied to
computer science, although the reader will probably see things like
\(A\limp B\) or \(A\Rightarrow B\) when we would write \(A\lneg \lpar
B\).

It is standard to say that one map is the \dfn{transpose} of the other
(an even more standard term is ``exponential transpose'' but in linear
logic the first adjective is dropped, for obvious reasons). The map
\(g\) is also often called the \dfn{curryfication} of \(f\). We will
call the reverse process \dfn{de-curryfication} and we will use
the term ``transpose'' in a generic, non-directional way.

We can apply the symmetry isomorphism to the evaluation map, and get a
map \(B\lneg\ltens (B\lpar C) \to C\), which is also ``the''
evaluation map. Both versions of the evaluation map will be simply
denoted by \(\epsilon_{B}\), since there is little chance of confusion.
In the same way, we allow ourselves to write the fundamental natural
bijection as \(\Hom_{\CC}(A,C\lpar B) \isom \Hom_{\CC}(C\lneg \ltens
A, B)\). This is obtained by using symmetry, but it could be true even
if we didn't have the symmetry, for example in the case of a cyclic
*-autonomous category~\cite{barr:95,blute:etal:02}. Thus
we can say we have ``left curryfying'' and ``right curryfying''.

\begin{proposition}\label{prop:currycomp}
  Let \(f\colon A\ltens B\lneg \to C\) and \(g\colon A \to C\lpar B\)
  be just as above, and let \(h\colon A'\to A\), \(k\colon B\to B'\),
  \(l\colon C \to C'\), \(m\colon B''\to B\lneg\).  Then:
  \begin{enumerate}
  \item\label{item:currycompone} the curryfication of \(f\circ (h
    \ltens m )\colon A'\ltens B^{\prime\prime} \to C\) is
    \((\id_{C} \lpar m\lneg )\circ  g\circ h\colon A'\to C\lpar
    B^{\prime\prime \perp}\),
  \item\label{item:currycomptwo} the curryfication of \(f\circ (h
    \ltens \id_{B\lneg})\colon A'\ltens B\lneg \to C\) is \(g\circ
    h\colon A'\to C\lpar B\),
  \item\label{item:currycompthree} the de-curryfication of \((l
    \lpar k) \circ g\colon A\to C' \lpar B'\) is \(l \circ f\circ
    (\id_{A} \ltens k\lneg)\colon A\ltens B^{\prime \bot}\to C'\),
  \item\label{item:currycompfour} the de-curryfication of
    \((l \lpar \id_B) \circ g\colon A\to C' \lpar B \) is 
    \(l\circ f\colon A\ltens B\lneg\to C' \).
  \end{enumerate}
\end{proposition}

\proof
  We can observe that \eqref{item:currycompone} is just a
  restatement of the naturality of the defining natural isomorphism of
  *-autonomous categories. Then \eqref{item:currycomptwo} 
  is obtained by replacing \(m\) by \(\id_{B\lneg}\). Or we first
  could prove \eqref{item:currycomptwo} by applying
  Proposition~\ref{prop:basiccurry} to \(g\circ h\) and seeing that it
  gives us exactly \(f\circ (h \ltens \id_{B\lneg})\), and then apply
  that proposition again, using duality and some exchange of left and
  right.
  
  The last two statements are just the duals of the first two.\qed

\begin{proposition}\label{prop:evalcomm}
  The following diagram always commutes:
  \[
  \xymatrix@C=4em@R=5ex{ A\lneg \ltens (A\lpar B\lpar C) \ltens C\lneg
    \ar[r]^-{\id_{A\lneg}\ltens \epsilon_{C}} &
    A\lneg \ltens (A\lpar B) \ar[d]^{\epsilon_{A}} \\
    (B\lpar C)\ltens C\lneg \ar@{<-}[u]^{\epsilon_{A}\ltens \id_{C
        \lneg}} \ar[r]_{\epsilon_{C}} & B }
  \]
\end{proposition}

\proof
  The previous proposition tells us that if we right- (or \(C\)-)
  curryfy \(\epsilon_{C} \circ (\epsilon_{A}\ltens \id_{C\lneg})\) we get
  \[
  \xymatrix@C=4em@R=5ex{
    A\lneg \ltens (A\lpar B\lpar C) \ar[r]^-{\epsilon_{A}} & B\lpar
    C\ar[r]^{\epsilont} & B\lpar C}\quadcm
  \]
  where the map \(\epsilont\) is the curryfication of
  \(\epsilon_C\colon (B\lpar C)\ltens C\lneg \to B \), i.e., the identity
  \(\id_{B\lpar C}\). If we then \(A\)-curryfy this composite, which
  is just \(\epsilon_{A}\) we get the identity on \(A\lpar B\lpar C\).
  We can do the same to \(\epsilon_{A}\circ (\id_{A\lneg}\ltens
  \epsilon_{C})\), this time \(A\)-curryfying before we \(C\)-curryfy,
  and we will also get the identity on \(A\lpar B \lpar C\). Thus the
  square commutes by uniqueness of transposes.  \qed
We could interpret this by saying that the operations of left and
right curryfying commute with each other. But because we have a
symmetry, it is better to say that any two successive applications of
curryfication will commute with each other, the left-right distinction
being purely for readability.

By curryfying the isomorphism \(\rho_{A} \colon A\ltens \lone \to A \)
we get an arrow \(\idh_{A}\colon \lone \to A\lneg \lpar A\), which is
often called the \dfn{name of the identity}. Its dual is
\(\idh_{A}\lneg \colon A\lneg \ltens A \to \lbot\). In general any map
\(f\colon A\to B\) has a \dfn{name} \(\hat f\colon \lone \to A\lneg
\lpar B\), obtained by curryfying \(f\circ \rho_{A} \colon A\ltens
\lone \to B\).

\begin{proposition}\label{prop:multi}
  Let $\CC$ be a *-autonomous category, and let $C_1,\ldots,C_n$ be
  objects of $\CC$. Let $I,J\subseteq\set{1,\ldots,n}$, and let 
  $\complI=\set{1,\ldots,n}\setminus I$ and
  $\complJ=\set{1,\ldots,n}\setminus J$ be their complements.
  Then for all bracketings of  $I$, $J$, $\complI$, and $\complJ$, 
  we have a natural bijection between 
  $$
  \Hom_\CC\left(\,\lbigtensind{I}{i}{C\i\lneg}\,,
  \,\lbigparind{\complI}{i}{C\i}\,\right)$$
  and
  $$\Hom_\CC\left(\,\lbigtensind{J}{j}{C\i\lneg}\,,
  \,\lbigparind{\complJ}{j}{C\i}\,\right)\quadfs$$
\end{proposition}

\proof
  This should be obvious in view of the previous results: one map is
  always obtained from the other by applying the ``transpose'' operator as
  needed.  \qed

In other words, any
$$f:\lbigtensind{I}{i}{C\i\lneg}\to\lbigparind{\complI}{i}{C\i}$$
uniquely determines an arrow
$$f':\lbigtensind{J}{j}{C\i\lneg}\to\lbigparind{\complJ}{j}{C\i}\quadcm$$
and vice versa.

This means that every
$$f:\lbigtensind{I}{i}{C\i\lneg}\to\lbigparind{\complI}{i}{C\i}$$
uniquely determines a whole family of arrows, indexed by the set of
bracketings on \(I\). We will call such a family an \dfn{equivariant
  family}, and a member of it a \dfn{representative}. Note that if we
put \(I = \emptyset\) we get representatives that are more canonical
than others, with source~\(\lone\); this is also the case when we put
\(\complI = \emptyset\), where the representatives have
target~\(\lbot\).  Proposition~\ref{prop:multi} ensures that every
representative of such a family of morphisms uniquely determines the
whole family.  This will turn our to be very helpful for the
construction of the free *-autonomous category in
Section~\ref{sec:freestar}; it allows us to avoid the notion of poly-
or multicategory \cite{lambek:69,szabo:75,cockett:seely:97}.

For example, any map, say \(f\colon A\to B\) will have (at least)
six members in
its equivariant family. There will be \(f\), with \( f\lneg \colon
B\lneg \to A\lneg\), the name \(\fh \colon \lone \to A\lneg \lpar B\)
and its dual \(\fh\lneg \colon B\lneg \ltens A \to \lbot \), along
with the twisted versions $\lone \to B\lpar A\lneg$ and $A \ltens B\lneg
\to \lbot$ of $\fh$ and $\fh\lneg$, respectively. 
These two do not deserve their own
special notation, and we will sometimes call them $\fh$ and $\fh\lneg$,
even if something like \(\sigma_{A,B} \circ \fh\) is really the
correct notation.

\bigskip

Let us give some more standard constructions on *-autonomous categories
and their relatives. Given arbitrary objects $A$, $B$, $C$, and $D$, 
take the tensor
\begin{equation}\label{eqn:intertens}
\vcenter{
\xymatrix@C=4em@R=5ex{ 
  A\lneg \ltens (A\lpar B)\ltens (C\lpar D) \ltens D\lneg
  \ar[r]^-{\epsilon_{A}\ltens\epsilon_{D}} & B\ltens C
}}
\end{equation}
and then curryfy twice, left and right. We get a natural\footnote{This
  fact will not be used afterwards and we won't prove it.} map
$$\myst_{A,B,C,D}\colon (A\lpar B)\ltens(C\lpar D)\longrightarrow
A\lpar (B\ltens C)\lpar D
\quadcm
$$
 the \dfn{internal tensor}. A particular
case of this is when \(A=\lbot\). Thus we can form
{\small
\[
\xymatrix@C=4em@R=5ex{ 
  B \ltens (C\lpar D) \ar[r]^-{\lambda^{-1}_{B}\ltens \id_{C\lpar
      D}}& (\lbot \lpar B) \ltens (C\lpar D)
  \ar[r]^-{\tau_{\lbot,B,C,D}}& \lbot \lpar( B \ltens C)\lpar D
  \ar[r]^-{\lambda_{(B\ltens C) \lpar D}}
  & (B\ltens C) \lpar D }
\]%
}%
and get an arrow, that we call
\dfn{switch}~\cite{guglielmi:strassburger:01,brunnler:tiu:01} but is
more traditionally known as \dfn{weak
  distributivity}~\cite{hyland:depaiva:fill,cockett:seely:97} or \dfn{linear
  distributivity}\footnote{We would like to add that this law is much
  more an artifact of associative logics than a form of
  distributivity, and that Do\v sen's coinage
  \emph{dissociativity}~\cite{dosen:petric:coherence-book,dosen:petric:05} for it should
  be considered seriously.}, and
that we denote by \(\tau_{\emptyset,B,C,D}\). There is another version
of switch, \(\tau_{A,B,C,\emptyset}\colon (A\lpar B )\ltens C \to
A\lpar (B\ltens C)\) obtained by replacing \(D\) by \(\lbot\).  An
interesting property of switch is that it is self-dual, i.e., 
\begin{equation}\label{eqn:switchduality}
    \tau\lneg_{\emptyset,B,C,D} = \tau_{\emptyset,D\lneg,C\lneg,B\lneg}
    \qquand
    \tau\lneg_{A,B,C,\emptyset} =
    \tau_{C\lneg,B\lneg,A\lneg,\emptyset} 
\end{equation}
as the reader can show.

\begin{proposition}\label{prop:namestensor}
  Let \(f\colon A\to B\) and \(g\colon D \to C\). Then the following holds:
  \begin{equation}\label{eqn:namestensor}
    \vcenter{
      \xymatrix@C=4em@R=5ex{
        \lone \ltens\lone
        \ar[r]^{\fh\ltens\gh\qquad\quad}
        \ar[dd]_{\isom} &
        (A\lneg\lpar B)\ltens(C\lpar D\lneg)
        \ar[d]^{\tau_{A\lneg,B,C,D\lneg}}\\
        {}&
        A\lneg\lpar(B\ltens C)\lpar D\lneg \ar[d]^{\isom} \\
        \lone
        \ar[r]_-{\widehat{f\ltens g}}& (D\lneg\lpar A\lneg)\lpar (B\ltens C) 
      }}
  \end{equation}
\end{proposition}
\proof
  Do left-right de-curryfication on \(\tau_{A\lneg ,B,C,D\lneg}
  \circ ( \hat f \ltens \hat g) \). We get a map \((A \ltens \lone)
  \ltens (\lone \ltens D) \to B\ltens C \)
  and we can precompose it with \(\rho^{-1}_{A}\ltens \lambda^{-1}_{D}\):
  {\small
  \[
  \qlapm{
    \xymatrix@C=1.4em@R=3ex{ A\ltens D  \ar[rd]^-{\quad\rho^{-1}_{A}\ltens
      \lambda^{-1}_{D}} \ar@{<->}[dd]_{\isom} & \\
    {} & (A\ltens \lone) \ltens (\lone \ltens
    D)\ar[rrr]^-{\id_{A}\ltens \hat f \ltens \hat g \ltens \id_{D}} &{}&&
    A\ltens (A\lneg\lpar B) \ltens (C\lneg \lpar D)\ltens D\lneg
    \ar[rr]^-{\epsilon_{A\lneg} \ltens \epsilon_{D}} && B\ltens C  \\
    A\ltens \lone\ltens D 
    \ar[ur]_-{\quad\id_{A}\ltens \delta \ltens \id_{D}} &
     }}
  \]
  }%
  (\(\delta\) being the obvious \(\rho_{\lone}^{-1} =
  \lambda_{\lone}^{-1}\)). It should be clear that this is
  \(\xymatrix@1{A\ltens D \ar[r]^-{f\ltens g}& B\ltens C } \). The
  bottom part of the triangle adds details on how the name
  \(\widehat{g\ltens f}\) fits in
  equation~(\ref{eqn:namestensor}).\qed
This gives an explanation for the name internal tensor. The following
is actually more general, but we will let the reader check that.

\begin{proposition}\label{prop:currytau}
  Let \(f\colon X\ltens Y \to A\lpar B\) and \(g\colon Z\ltens W \to C
  \lpar D\) be maps, and let \(\tilde f\colon Y \to X\lneg \lpar
  A\lpar B\) and \( \tilde g\colon Z\to C \lpar D\lpar W\lneg \) be their
  curryfications. Then the left-right curryfication of 
  \[
  \xymatrix@C=4em@R=3ex{ 
    X\ltens Y \ltens Z \ltens W \ar[r]^-{ f \ltens g} &
    (A\lpar B)\ltens (C\lpar D) \ar[r]^-{\tau_{A,B,C,D}} & A\lpar
    (B\ltens C) \lpar D
  }
  \]
  is
  {\small
  \[
  \xymatrix@C=2em@R=3ex{
    Y\ltens Z \ar[r]^-{\tilde f \ltens
      \tilde g} & \bigl( X\lneg\lpar A \lpar B\bigr) \ltens \bigl(C
    \lpar D \lpar W\lneg \bigr) \ar[rrr]^-{\tau_{X\lneg \lpar A, B, C,
        D\lpar W\lneg}} &&& X\lneg \lpar A \lpar (B\ltens C) \lpar D
    \lpar W\lneg }\,.
  \]
  }
\end{proposition}

\proof
  If we de-curryfy the first map on \(A, D\) we get (by using
  Proposition~\ref{prop:basiccurry} twice) 
  \[
  \xymatrix{
    A\lneg \ltens X\ltens Y \ltens Z\ltens W \ltens D\lneg
    \ar[d]^{\id_{A}\ltens f\ltens g \ltens \id_{D}} \\
    A\lneg \ltens (A\lpar B) \ltens (C\lpar D) \ltens D\lneg
    \ar[d]^{\epsilon_{A} \ltens \epsilon_{B}} \\
    B\ltens C
    }
  \]
  while if we de-curryfy the second map on \(X\lneg \lpar A\) and
  \(D\lpar W\lneg\) we get
  \[
  \xymatrix{
    (A\lneg \ltens X) \ltens Y \ltens Z \ltens (W \ltens D\lneg)
    \ar[d]^{\id_{A\ltens X\lneg} \ltens \tilde f \ltens \tilde g \ltens \id_{W \ltens D\lneg}} \\
    (A\lneg \ltens X) \ltens (X\lneg \lpar A \lpar B) \ltens (C\lpar D\lpar W\lneg) \ltens (W\ltens D\lneg)
    \ar[d]^{\epsilon_{X\lneg \lpar A}\ltens \epsilon_{D \lpar W\lneg}} \\
    B\ltens C
}
  \]
  But these two are equal: just apply Proposition~\ref{prop:evalcomm}
  (with the remark right after it) twice, along with the defining
  universal property of the transpose operation, i.e.,
  Proposition~\ref{prop:basiccurry}.\qed

\begin{proposition}\label{prop:twotausareone}
  The following commutes
  \[
  \xymatrix{ (A \lpar B) \ltens (C\lpar D) \ar[rr]^-{\tau_{A,B,C,D}}
    \ar[rd]_-{\tau_{\emptyset,A\lpar B, C, D}\quad} &&
    A\lpar (B\ltens C) \lpar D  \\
    & \bigl((A\lpar B ) \ltens C\bigr) \lpar D
    \ar[ru]_-{\quad\tau_{A,B,C,\emptyset}\lpar\id_{D}}\,\,.  & }
  \]
\end{proposition}
\proof
  We know we have proved this if we can show that the left-right
  de-curryfication of the composite map gives us \(\epsilon_{A}
  \ltens \epsilon_{D}\); this is just by definition of \(\tau\). It
  should also be clear that the right de-curryfication of
  \(\tau_{\emptyset,A\lpar B, C, D}\) is \(\id_{A\lpar B} \ltens
  \epsilon_{D}\), and that the left de-curryfication of
  \(\tau_{A,B,C,\emptyset}\) is \(\epsilon_{A} \ltens \id_{C}\). From
  the first fact it follows that the right de-curryfication 
  of the composite map is
  \[
    \xymatrix@C=4em@R=3ex{
      (A\lpar B) \ltens (C\lpar D)\ltens D\lneg 
      \ar[r]^-{\id_{A\lpar B}\ltens\epsilon_{D}} & (A\lpar B)\ltens C
      \ar[r]^-{\tau_{A,B,C,\emptyset}}& A\lpar(B\ltens C)}
    \quadcm
  \]
  and using the second fact it is easy to see that the left
  de-curryfication of this is \(\epsilon_{A}\ltens \epsilon_{D} \).\qed

We can also construct the composite 
{\small
\[
\clapm{\xymatrix@C=2.6em@R=5ex{ (A\lpar B)\ltens(B\lneg \lpar
  C)\ar[rr]^-{\myst_{A,B,B\lneg,C}} && A\lpar (B\ltens B\lneg )\lpar C
  \ar[rr]^-{\id_{A\lneg}\lpar \idh\lneg_B\lpar \id_{C}}&&
  A\lneg\lpar \lbot \lpar C \ar[r]^-{\cong} & A\lneg \lpar C
}}
\]
}%
We call this arrow $\gamma_{A,B,C}:(A\lneg\lpar B)\ltens(B\lneg\lpar
C)\to A\lneg\lpar C$ the \dfn{internalized composition}. It should be
clear (apply Proposition~\ref{prop:currycomp} twice) that it
is obtained by applying left-right curryfication on
\[
\xymatrix@C=4em@R=5ex{ A\ltens (A\lneg \lpar B)\ltens (B\lneg \lpar C)\ltens
  C\lneg \ar[r]^-{\epsilon_{A\lneg}\ltens \epsilon _{C}} & B\ltens
  B\lneg \ar[r]^-{\idh\lneg_B} & \lbot
}
\]
The name internal composition can also be explained:

\begin{proposition}\label{prop:namescompose}
  Let $f\colon A\to B$ and $g\colon B\to C$.
  Then the following always commutes:
\begin{equation}\label{eqn:gamma}
\vcenter{
\xymatrix@C=4em@R=5ex{
\lone\ltens\lone
\ar[r]^{\fh\ltens\gh\qquad\quad}
\ar[d]_{\isom} &
(A\lneg\lpar B)\ltens(B\lneg\lpar C)
\ar[d]^{\gamma_{A,B,C}}\\
\lone
\ar[r]_-{\widehat{f\fcomp g}}&
A\lneg\lpar C 
}}
\end{equation}
\end{proposition}
\proof
  Repeat the proof of~\ref{prop:namestensor}, starting by a
  left-right de-curryfication on \(\gamma_{A,B,C}\circ ( \hat f \ltens
  \hat g) \), then doing precomposition with \(\rho^{-1}_{A}\ltens
  \lambda^{-1}_{C\lneg}\).  When reaching the sentence ``It should be
  clear that\ldots'', what should be clear now is that we are looking
  at
  \[
  \xymatrix@C=4em@R=5ex{A\ltens C\lneg \ar[r]^-{f\ltens g\lneg}&
    B\ltens B\lneg \ar[r]^-{\idh\lneg_B} & \lbot \,\,.}
  \]
  Seen as an
  equivariant family, one representative is \(g\circ f\) and another
  is \(\widehat{g\circ f}\). \qed

We have never seen the following in the literature. Perhaps it can be
considered trivial for a seasoned category theorist, but we think it
is worthwhile proving in full.

\begin{proposition}[Two-Tensor Lemma]\label{prop:twotens}
  The following always commutes:
  {\small
  \begin{equation}\label{eqn:twotenssqr}
    \vcenter{
    \xymatrix@C=3em@R=5ex{ (X\lpar A)\ltens (B\lpar Y\lpar C) \ltens (D\lpar Z)
    \ar[d]_{\myst_{X,A,B,Y\lpar C} \ltens \id_{D\lpar
        Z}}\ar[rr]^{\id_{X\lpar A} \ltens \myst_{B\lpar Y, C, D, Z}} &\quad &
  (X\lpar A) \ltens \bigl( B \lpar Y \lpar (C\ltens D)\lpar Z \bigr)
  \ar[d]^{\myst_{X, A, B, Y\lpar(C\ltens D)\lpar Z}} \\
  {}\bigl(X\lpar (A\ltens B) \lpar Y \lpar C \bigr)\ltens (D\lpar Z)
  \ar[rr]_{\myst_{X\lpar (A\ltens B)\lpar Y,C,D,Z}} &\quad & X\lpar (A\ltens B)
  \lpar Y \lpar (C\ltens D)\lpar Z
  }}
  \end{equation}
  }
\end{proposition}
\proof
  Let \(M = X\lneg \ltens (X \lpar A)\ltens (A\lneg \lpar B\lneg )\)
  and \(N = (C\lneg \lpar D\lneg )\ltens (D\lpar Z) \ltens Z\lneg\).
  There are obvious \(m\colon M\to B\lneg \) and \(n\colon N \to
  C\lneg\), which are just sequences of evaluations. We want to show
  that the two ways of computing the diagonal above are equal. If we
  de-curryfy these maps left and right enough times, they both
  can be considered as maps:
  \[
  X\lneg \ltens (A\ltens B)\lneg \ltens (X\lpar A)\ltens (B\lpar
  Y\lpar C) \ltens (D\lpar Z) \ltens (C\ltens D)\lneg \ltens Z\lneg
  \quad\longrightarrow\quad Y
  \]
  whose source is isomorphic to \(M\ltens (B\lpar Y\lpar C) \ltens N
  \), modulo some symmetries.
  Now look at
  {\small
  \begin{equation}\label{eqn:twotensorprf}
    \vcenter{
    \xymatrix{ 
      M\ltens (B\lpar Y\lpar C) \ltens N \ar[rr]^{\id_{M}\ltens \id
      \ltens n} \ar[d]_{ m \ltens \id \ltens \id_{N}} &\quad & M\ltens
    (B\lpar Y\lpar C) \ltens C\lneg \ar[d]_{m\ltens \id\ltens \id_{C\lneg}
    } \ar[r]^-{\id_{M}\ltens \epsilon_{C}} &
    M\ltens (B\lpar Y) \ar[d]^{m\ltens \id_{B\lpar Y}}    \\
    B\lneg \ltens (B\lpar Y\lpar C) \ltens N
    \ar[d]_{\epsilon_{B}\ltens \id_{N}}\ar[rr]_{\id_{B\lneg} \ltens \id
        \ltens n} &\quad & B\lneg \ltens (B\lpar Y\lpar C) \ltens
    C\lneg \ar[r]_-{\id_{B\lneg} \ltens
      \epsilon_{C}}\ar[d]_{\epsilon_{B} \ltens \id_{C\lneg}} &
    B\lneg \ltens (B\lpar Y) \ar[d]^{\epsilon_B} \\
    (Y\lpar C) \ltens N \ar[rr]_{\id_{Y\lpar C}\ltens n} &{}& (Y\lpar
    C)\ltens C\lneg \ar[r]_{\epsilon_{C}} & Y
  }}
  \end{equation}%
  }%
  The only small square in there that does not commute trivially is
  the bottom right one, and it commutes because
  of~Proposition~\ref{prop:evalcomm}. But compare the outer square
  above with the previous one. Take one path
  of~(\ref{eqn:twotenssqr}), say, first right, then down. We get a map
  of the form \(\myst \circ ( \id\ltens \myst)\). If we curryfy it
  twice, we get exactly the corresponding (right-right-down-down) path
  in~(\ref{eqn:twotensorprf}). The same argument applies to the
  down-down-right-right path, and then since~(\ref{eqn:twotensorprf}) 
  commutes we
  get the result by uniqueness of transposes.  \qed

\subsection{Proof nets form a *-autonomous category}\label{sec:PNstar}

The first basic observation of this section is that the proof nets
that we have defined in Section~\ref{sec:pn}
form a category. For making this precise, we
provide for every formula $A$ an identity proof net
$\id_A=\prfnet{I_A}{A\lneg,A}$, where $I_A$ is called the
\dfn{identity linking} which is defined inductively on $A$ as follows:
$$
\begin{array}{lllll}
I_a&=&I_{a\lneg}&=&a\ltens a\lneg\\
I_\lbot&=&I_\lone&=&\lbot\ltens\lone\\
I_{A\lpar B}&=&I_{A\ltens B}&=&I_A\lpar I_B
\end{array}
$$
Observe that we can have that $I_A=I_{A\lneg}$ because changing the
order of the arguments of a $\ltens$ or $\lpar$ in the linking of a
proof graph does not change the proof net (see~\ref{par:grfequ}).

Furthermore, for any two proof nets $f=\prfnet{P}{A\lneg,B}$ and
$g=\prfnet{Q}{B\lneg,C}$, we can define their composition $g\fcomp f$ 
to be the result of applying the cut
elimination procedure to $\prfnet{P\lpar Q}{A\lneg,B\lcut B\lneg,C}$.
That this is well-defined and associative follows immediately from the
strong normalization of cut elimination. We also have that
$f\fcomp\id_A=f=\id_B\fcomp f$. 

This gives rise to a category $\PN(\cA)$ whose objects are the
formulas built from $\cA\cup\cA\lneg\cup\set{\lbot,\lone}$ via
$\ltens$ and $\lpar$ (cf.~\eqref{eq:formula} on page
\pageref{eq:formula}), and whose arrows are the proof nets. More
precisely, the arrows between two objects $A$ and $B$ are the
(cut-free) proof nets $\prfnet{P}{A\lneg,B}$ (see
Definition~\ref{def:proofnet} on page~\pageref{def:proofnet}).

The main result of this section is the following:

\begin{proposition}
  For every set $\cA$, the category $\PN(\cA)$ is a (strict)
  *-autonomous category.  
\end{proposition}

\proof
  The unit object is given by the formula $\lone$, and the bifunctor
  $-\ltens-:\PN(\cA)\times\PN(\cA)\to\PN(\cA)$ is determined by the operation  
  $\ltens$ on formulas, because for any two proof nets
  $f=\prfnet{P}{A\lneg,B}$ and $g=\prfnet{Q}{C\lneg,D}$
  we have the proof net
  $f\ltens g = \prfnet{P\lpar Q}{C\lneg\lpar A\lneg,B\ltens D}$.
  We can exhibit the natural isomorphisms $\alpha$, $\sigma$, $\rho$ and
  $\lambda$, which are required by the definition of symmetric monoidal
  categories as follows:
  $$
  \clap{$\begin{array}{r@{\;=\;}l@{\;\colon\;}l}
      \alpha_{A,B,C}&\prfnet{I_A\lpar I_B\lpar I_C}%
            {(C\lneg\lpar B\lneg)\lpar A\lneg,(A\ltens B)\ltens C}&
            A\ltens(B\ltens C)\to(A\ltens B)\ltens C\\
            \rho_A&\prfnet{\lbot\ltens I_A}{\lbot\lpar A\lneg,A}&
            A\ltens\lone\to A\\
            \lambda_A&\prfnet{\lbot\ltens I_A}{A\lneg\lpar\lbot,A}&
            \lone\ltens A\to A\\
            \sigma_{A,B}&\prfnet{I_A\lpar I_B}%
                  {B\lneg\lpar A\lneg,B\ltens A}&
                  A\ltens B\to B\ltens A
    \end{array}
    $}
  $$
  It is easy to check that these are indeed proof nets, that they
  are natural isomorphisms, and that the diagrams given in Definitions
  \ref{def:monoidal} and~\ref{def:symmetric} do indeed commute.  The
  duality functor $\fneg$ is defined on the objects as in
  \eqref{eq:neg} on page~\pageref{eq:neg}, and on arrows by assigning
  to $f=\prfnet{P}{A\lneg,B}:A\to B$ the arrow
  $f\lneg=\prfnet{P}{B,A\lneg}:B\lneg\to A\lneg$.  Observe that in
  this particular case we have that $A\lnegneg\to A$ is the identity,
  and not just an isomorphism. This will be discussed in detail in the
  next section. For now, it only remains to check that we have our
  natural bijection
  \begin{eqnarray*}
    \Hom(A\ltens B,C)&\isom&\Hom(B,A\lneg\lpar C)\\
    \prfnet{P}{B\lneg\lpar A\lneg,C}&\mapsto&\prfnet{P}{B\lneg,A\lneg\lpar C}
    \quadfs
  \end{eqnarray*}
  That we have $A\lpar B=(B\lneg\ltens A\lneg)\lneg$ and $\lbot=\lone\lneg$
  does not come as a surprise.
  \qed

\subsection{The free *-autonomous category}\label{sec:freestar}

In this section we will show that the category of proof nets is the
free strict *-autonomous category: we have already observed that our
\(\PN(\cA)\) is strict, in the sense that $A\to A\lnegneg$ is always
the identity for every object $A$. So let $\cA$ be any set and let
$\eta_\cA:\cA\to \Obj(\PN(\cA))$ be the function that maps every
element of $\cA$ to itself seen as atomic formula.  To say that
$\PN(\cA)$ is the \dfn{free (strict) *-autonomous category generated
  by $\cA$} amounts to saying that

\begin{theorem}\label{thm:freestar}
Given a strict *-autonomous category 
$(\CC,\ltens,\lone_\CC,\fneg)$ and a
map $\Gnull:\cA\to\Obj(\CC)$, there is a unique functor
 $G:\PN(\cA)\to\CC$, preserving the *-autonomous structure,  
such that $\Gnull=\Obj(G)\fcomp\eta_\cA$,
where $\Obj(G)$ is the restriction of $G$ on objects.
\end{theorem}

The remainder of this section is devoted to the proof of this
theorem.  

Let the *-autonomous category $\CC$ and the embedding
$\Gnull:\cA\to\Obj(\CC)$ be given. We will exhibit the functor 
$G:\PN(\cA)\to\CC$ which has the desired properties. On the objects,
this functor is uniquely determined as follows:
$$
\clapm{\begin{array}{rcl@{\quad\;}rcl@{\quad\;}rcl}
G(a)&=&\Gnull(a)&G(\lbot)&=&\lbotC&G(A\lpar B)&=&G(A)\lpar G(B)\\
G(a\lneg)&=&\Gnull(a)\lneg&G(\lone)&=&\loneC&G(A\ltens B)&=&G(A)\ltens G(B)\\
\end{array}}
$$
There is no other choice, since the objects $\loneC$ and $\lbotC$
along with the functors $(-)\lneg$, $-\ltens-$, and $-\lpar-$ are
uniquely determined by the *-autonomous structure on $\CC$.

For defining $G$ on the morphisms, the situation is not as simple.  In
fact, before saying how $G$ acts on proof nets, we will first define a
mapping $\Gflat$ that assigns to each ordinary proof net with cuts
(see \ref{par:unitfree} and \ref{rem:unitfreecut}) an equivariant
family of arrows in $\CC$ (as defined in Section~\ref{sec:defstar}).
More precisely, let $\pi$ be an ordinary proof net with conclusions
$A_0,\ldots,A_n,B_1\lcut B\lneg_1,\ldots,B_m\lcut B\lneg_m$ (for some
$n,m\ge 0$), where $A_0,\ldots,A_n$ are the formulas in the sequent
that are not cuts, and $B_1\lcut B\lneg_1,\ldots,B_m\lcut B\lneg_m$
are the cuts.  For $\pi$ we will construct a uniquely defined
equivariant family $\Gflat(\pi)$ of arrows
$$
\lbigtensind{I}{i}{G(A\i)\lneg}\to
\lbigparind{\complI}{i}{G(A\i)}
$$
indexed by the bracketings on the subsets
$I\subseteq\set{0,\ldots,n}$ and their complements.  We begin with the
cut-free case, \ie the case where $m=0$.  We proceed by induction on
the size of $\pi$ (\ie the sum of the numbers of $\ltens$- and
$\lpar$-nodes).  We again make crucial use of
Lemma~\ref{lem:splitting}, the existence of a splitting tensor.
\begin{itemize}
\item If the net contains no $\ltens$- or $\lpar$-nodes, 
then it is a single ordinary axiom link with
conclusion $a,a\lneg$. In this case our equivariant family is 
determined by the identity $\id:G(a)\to G(a)$. 
\item If one of the root nodes in the net is a $\lpar$, \ie
  $A_j=A'_j\lpar A''_j$ for some $j\in\set{0,\ldots,n}$, then we have
  by induction hypothesis the 
  equivariant family with representative
  $$
  \lbigtens{}%
  {\bigset{G(A_i)\lneg\mid i\in\set{0,\ldots,n}\setminus\set{j}}}\to
  G(A'_j)\lpar G(A''_j)
  $$
  from which we get immediately
  $$
  \lbigtens{}%
  {\bigset{G(A_i)\lneg\mid i\in\set{0,\ldots,n}\setminus\set{j}}}\to
  G(A_j)
  $$
  because $G(A'_j)\lpar G(A''_j)=G(A_j)$. 
\item If one of the roots is a splitting $\ltens$, say $A_j=A'_j\ltens
  A''_j$, then by removing the $\ltens$-root we can get two smaller
  ordinary proof nets $\pi_1$ and $\pi_2$, which are both correct.
  \Wolg, $\pi_1$ has conclusions $A_0,\ldots,A_{j-1},A'_j$ and $\pi_2$
  has conclusions $A''_j,A_{j+1},\ldots,A_n$ (\ie we might have to
  choose a different ordering
  of the $A_i$).  By induction hypothesis, we have two 
  equivariant families $\Gflat(\pi_1)$ and $\Gflat(\pi_2)$, with
  representatives
  $$
  \lbigtens{}{\bigset{G(A_0)\lneg,\ldots,G(A_{j-1})\lneg}}\to G(A'_j)\quand
  $$
  $$
  \lbigtens{}{\bigset{G(A_{j+1})\lneg,\ldots,G(A_n)\lneg}}\to G(A''_j)
  \quadcm \phantom{\quand}      
  $$
  respectively,
    from which we get
    $$
    \lbigtens{}{\bigset{G(A_i)\lneg|i\in\set{0,\ldots,n}\setminus\set{j}}}\to 
    G(A_j)
    $$
    by applying the functor $-\ltens-$ and the fact that
    $G(A_j)=G(A'_j)\ltens G(A''_j)$. 
\end{itemize}
In all three cases the construction is uniquely determined by the *-autonomous
structure on $\CC$ and the choice of atoms. 

\begin{remark}\label{rmrk:tensorintro}
  Let $\pi_{1}$ and $\pi_{2}$ be as in the last case.  We can choose
  representatives \(r\colon \lone \to G( A_{0}) \lpar
  \cdots \lpar G(A_{j-1}) \lpar G(A_{j}^{\prime})\) and \(s\colon
  \lone \to G(A_{j}^{\prime\prime}) \lpar G(A_{j+1})\lpar \cdots \lpar
  G(A_{n}) \) for \(\Gflat(\pi_{1})\) and \( \Gflat(\pi_{2})\). Then
  because of Proposition~\ref{prop:namestensor} the following
  $$
  \xymatrix{ \lone \ar[d]^*+{\isom} \\
    \lone \ltens \lone \ar[d]^*+{\rlapm{r\ltens s}}\\
    \qqqlapm{\bigl(G( A_{0}) \lpar \cdots \lpar G(A_{j-1}) \lpar
    G(A_{j}^{\prime})\bigr) \ltens \bigl(G(A_{j}^{\prime\prime}) \lpar
    G(A_{j+1})\lpar \cdots \lpar G(A_{n})\bigr)}
    \ar[d]^*+{\rlapm{\tau_{G(A_0\cdots
        A_{j-1}),G(A'_{j}),G(A''_{j}),G(A_{j+1}\cdots A_n) }}} \\
    \qlapm{G(A_{0})\lpar \cdots \lpar G(A_{j-1}) \lpar G(A_{j}) \lpar
    G(A_{j+1}) \lpar \cdots \lpar G(A_{n})} }
  $$
  is a representative of \(\Gflat(\pi)\). We will also need a more
  general version of this: let \(\{0,\ldots,j-1 \}= L \cup L'\) and
  \(\{j+1,\ldots,n\} = K\cup K'\) be partitions in arbitrary disjoint
  subsets, and choose bracketings on $L$, $L'$, $K$, and $K'$. Let
  $$
  r'\colon
  \lbigtens{L'}\bigset{G(A_{l})\lneg \mid l\in L'} \longrightarrow
  \lbigpar{L}\bigset{G(A_{l}) \mid l\in L} \lpar G(A'_{j})
  $$
  and
  $$
  s'\colon
  \lbigtens{K'}\bigset{G(A_{k})\lneg \mid k\in K'} \longrightarrow
  G(A^{\prime\prime}_{j})\lpar \lbigpar{K}\bigset{G(A_{k}) \mid k\in K} 
  $$
  be representatives of $\Gflat(\pi_{1})$ and $\Gflat(\pi_{2})$
  respectively. Then it should be clear, because of
  Proposition~\ref{prop:currytau}, that
  \[
  \xymatrix{ \bigl(\lbigtens{L'}\set{G(A_{l})\lneg \mid l\in L}\bigr)
    \ltens \bigl(\lbigtens{K'}\set{G(A_{k})\lneg \mid k\in K'}\bigr)
    \ar[d]^*+{r'\ltens s'} \\
    \bigl(\lbigpar{L}\set{G(A_{l}) \mid l\in L} \lpar G(A'_{j})\bigr)
    \ltens \bigl(G(A''_{j})\lpar \lbigpar{K}\set{G(A_{k}) \mid k\in
      K}\bigr) \ar[d]^*+{\tau_{\lpar_{L}\set{G(A_{l}) \mid l\in
          L},G(A'_{j}),G(A''_{j}),
          \lpar_{K}\set{G(A_{k}) \mid k\in K}}} \\
    \lbigpar{L}\set{G(A_{l}) \mid l\in L} \lpar G(A_{j}) \lpar
    \lbigpar{K}\set{G(A_{k}) \mid k\in K} }
  \]
  is a representative of \(\pi\).
\end{remark}
It remains to show that the construction is independent from the order
in which the $\lpar$- and $\ltens$-nodes are introduced.  We will again
proceed by induction on the number of $\lpar$- and $\ltens$-nodes in
the net. In the base case, where there are fewer than two such nodes
in the net, we have uniqueness immediately. For the inductive case,
consider the last two nodes that have been introduced. If one of them
is not a root, then the other is its parent, and there is no
possibility in changing the order of the introductions. So assume both
of them are roots.  There are three cases to consider:
\begin{itemize}
\item Both of them are $\lpar$-nodes, say $A_j=A'_j\lpar A''_j$ and
  $A_k=A'_k\lpar A''_k$ for some $j,k\in\set{0,\ldots,n}$. Then we
  have by induction hypothesis the unique equivariant family with
  representative
  $$
  \lbigtens{} {\bigset{G(A_i)\lneg\mid
      i\in\set{0,\ldots,n}\setminus\set{j,k}}}\to
  \lbigpar{}{\bigset{G(A'_j),G(A''_j),G(A'_k), G(A''_k)}}
  $$
  from which we immediately get
  $$
  \lbigtens{} {\bigset{G(A_i)\lneg\mid
      i\in\set{0,\ldots,n}\setminus\set{j}}}\to G(A_j)\lpar G(A_k)
  $$
  because $G(A'_j)\lpar G(A''_j)=G(A_j)$ and $G(A'_k)\lpar
  G(A''_k)=G(A_k)$.  Uniqueness follows immediately from the
  associativity of the functor $-\lpar-$.
\item One is a $\lpar$ and the other is a $\ltens$, say 
  $A_j=A'_j\lpar A''_j$ and $A_k=A'_k\ltens A''_k$ for some 
  $j,k\in\set{0,\ldots,n}$. Then the $\ltens$ must be splitting, and
  the formula $A'_j\lpar A''_j$ must belong to one of the two parts
  (if this is not the case, \ie $A'_j$ is in one part and $A''_j$ in
  the other, then the $\ltens$ must be introduced before the $\lpar$,
  and we have uniqueness immediately).
  \Wolg, assume now that $A'_j\lpar A''_j$ is in the part of
  $A'_k$, that $k=j+1$, that the formulas $A_0,\ldots,A_{j-1}$ are also
  in the part containing $A'_k$, and that the formulas
  $A_{k+1},\ldots,A_n$ are in the
  part containing $A''_k$.
  Then we have by induction hypothesis two unique
  equivariant families with representatives
  $$
  \lbigtens{}{\bigset{G(A_0)\lneg,\ldots,G(A_{j-1})\lneg}}\to 
  \lbigpar{}{\bigset{G(A'_j),G(A''_j),G(A'_k)}}\quand
  $$
  $$
  \lbigtens{}{\bigset{G(A_{k+1})\lneg,\ldots,G(A_n)\lneg}}\to
  G(A''_k) \quadcm \phantom{\quand}
  $$
  from which we get (by Remark~\ref{rmrk:tensorintro}) immediately
  the unique equivariant family
  $$
  \lbigtens{} {\bigset{G(A_i)\lneg\mid
      i\in\set{0,\ldots,n}\setminus\set{j,k}}}\to G(A_j)\lpar
  G(A_k)\quadfs
  $$
\item Both of them are $\ltens$-nodes, say $A_j=A'_j\ltens A''_j$ and
  $A_k=A'_k\ltens A''_k$ for some $j,k\in\set{0,\ldots,n}$. Then both
  of them must be splitting (otherwise they cannot have been
  introduced consecutively.)  \Wolg, we can decree that $j<k$, and
  that by removing the two $\ltens$-roots, we get three smaller nets,
  where the first contains the formulas $A_0,\ldots,A_{j-1},A'_j$, the
  second contains $A''_j,A_{j+1},\ldots,A_{k-1},A_k'$, and the third
  contains $A''_k,A_{k+1},\ldots,A_n$.  By induction hypothesis, we
  have three uniquely determined equivariant families, with
  representatives
  $$
  \begin{array}{rcl}
    \lone&\to&
    \lbigpar{}{\bigset{G(A_0),\ldots,G(A_{j-1}),G(A'_j)}}
    \quadcm\\[2ex]
    \lone&\to&
    \lbigpar{}{\bigset{G(A''_j),G(A_{j+1}),\ldots,G(A_{k-1}),G(A'_k)}}
    \rlap{\quad,\quand}\\[2ex]
    \lone&\to&
    \lbigpar{}{\bigset{G(A''_k),G(A_{k+1}),\ldots,G(A_n)}}\quadfs
  \end{array}
  $$
  There are now two ways of constructing the representative
  $$
  \begin{array}{rcl}
    \lone&\to&
    \lbigpar{}{\bigset{ G(A_0),\ldots,G(A_n)}}\quadfs
  \end{array}
  $$
  It follows immediately from the two-tensor lemma
  (Proposition~\ref{prop:twotens}), that both yield the same
  equivariant family.
\end{itemize}

We now extend the construction of the equivariant families to ordinary
proof nets with cuts, \ie the case where $m>0$.  This is done by first
replacing each cut $B_i\lcut B\lneg_i$ in $\pi$ by the
$\ltens$-formula $B_i\ltens B\lneg_i$, and then applying the previous
construction to the net with conclusions $A_0,A_1,\ldots,A_n,B_1\ltens
B\lneg_1,\ldots,B_m\ltens B\lneg_m$, which yields in particular the
representative
$$
\lbigtens{}{\bigset{G(A_i)\lneg\mid i\in\set{0,\ldots,n}}}\to 
\lbigpar{}{\bigset{G(B_j)\ltens G(B_j)\lneg\mid j\in\set{1,\ldots,m}}}\quadfs
$$
Look at the map $\idh\lneg_{G(B_j)}\colon G(B_j)\ltens
G(B_j)\lneg\to\lbot$ (``the co-name of the identity'') which exist for
every $B_j$.  By taking the par of the family
\((\idh\lneg_{G(B_{j})})\), we construct
$$
\zeta: 
\lbigpar{}{\bigset{G(B_j)\ltens G(B_j)\lneg\mid j\in\set{1,\ldots,m}}}
\;\to\;\lbigpar{}\bigset{\lbot \mid j\in\set{1,\ldots,m} } \isom\lbot
$$
and by composition we get
$$
\lbigtens{}{\bigset{G(A_i)\lneg\mid i\in\set{0,\ldots,n}}}\to\lbot\quadcm
$$
which we define as a representative of the equivariant
family $\Gflat(\pi)$.
The uniqueness of this follows from the
functoriality and associativity of $-\ltens-$.
\begin{remark}\label{rmrk:onecutelim}
  Suppose that we have nets \(\pi_{1}\) on \(\Gamma,A\) and
  \(\pi_{2}\) on \(A\lneg,\Delta\). Let \(f_{1}\colon \Gamma\lneg\to
  A\) and \(f_{2} \colon A \to \Delta\) be representatives of
  $\pi_{1}$ and $\pi_{2}$, respectively. It should be clear that if we
  construct \(\pi\) by cutting \(\pi_{1},\pi_{2}\) on \(A,A\lneg\),
  then \(f_{2}\circ f_{1}\) is a representative of \(G(\pi)\). This is
  obtained by looking at the definition of the internal composition
  \(\gamma\) as well as Proposition~\ref{prop:namescompose} and
  Remark~\ref{rmrk:tensorintro}.
\end{remark}
A crucial observation about this construction is that eliminating cuts
from an ordinary proof net (see~\ref{par:unitfreecutelim}) does not
affect the equivariant family it defines:

\begin{lemma}\label{lem:unitfree}
  Let $\pi$ be an ordinary proof net with conclusions 
  $$A_0,\ldots,A_n,B_1\lcut B\lneg_1,\ldots,B_m\lcut B\lneg_m$$ 
  (for some $n,m\ge 0$), where $A_0,\ldots,A_n$ are the formulas in the
  sequent that are not cuts, and $B_1\lcut B\lneg_1,\ldots,B_m\lcut B\lneg_m$
  are the cuts. Let $\pi'$ be the ordinary proof net with conclusions
  $$ 
  A_0,\ldots,A_n\quadcm
  $$ 
  that is obtained from $\pi$ by applying the cut elimination
  procedure. Then $\pi$ and $\pi'$ determine the same equivariant family of
  arrows
  $$
  \lbigtensind{I}{i}{G(A\i)\lneg}\to
  \lbigparind{\complI}{i}{G(A\i)}
  $$ 
  indexed by the bracketings on the subsets $I\subseteq\set{0,\ldots,n}$
  and their complements.
\end{lemma}

\begin{remark}
  This lemma would suffice to prove that ``ordinary proof nets with
  two conclusions form the free *-autonomous category without units'',
  and is also an immediate consequence of this fact~\cite{blute:93}.
  But it is only recently that precise and fully satisfactory
  definitions for a notion of *-autonomous category without units have
  been proposed
  \cite{lam:str:05:freebool,houston:etal:unitless,dosen:petric:05},
  and we will not pursue this matter any further here.
\end{remark}

\def\mybreak{\hspace*{\fill} \cr \hspace*{\fill}}

\proof[Proof of Lemma~\ref{lem:unitfree}]
  The proof will be done by induction on the length of the cut
  reduction. It suffices to show the lemma for the case where $\pi'$
  is obtained from $\pi$ by a single cut reduction step. 
  We use the following convention: Given a sequent of \(n\)
  formulas \(\Gamma = A_{1},\ldots,A_{n}\) we write $G(\Gamma)$ for
  $G(A_1)\lpar\cdots\lpar G(A_k)$.
  We will now 
  proceed by induction on the size of $\pi$. There are four cases
  to consider:
  \begin{enumerate}
  \item If $\pi$ contains a $\lpar$-root, then this $\lpar$-root is
    also present in $\pi'$. Therefore we can remove it in both nets
    and apply the induction hypothesis.
  \item If $\pi$ is of the following shape:
    $$
    \vcenter
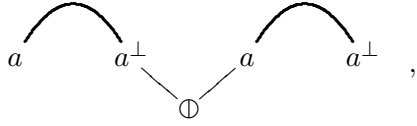
{
      \begin{xy}
        0;<2em,0ex>:<0em,4ex>::
        (0,1)="a"*{\sa}, (2,1)="b"*{\sna}, (4,1)="c"*{\sa},
        (6,1)="d"*{\sna}, 
        (3,0)="h"*{\lcut}, 
        "b"*{\ptens};"h"*{\ptens} **\dir{-},
        "c"*{\ptens};"h"*{\ptens} **\dir{-},
        "a"*{\ptens};"b"*{\ptens} **\crv{~*=<.5pt>{.}(1,\doubleheight)},
        "c"*{\ptens};"d"*{\ptens} **\crv{~*=<.5pt>{.}(5,\doubleheight)},
      \end{xy}
      }
    \quadcm
    $$
    \ie it consists of a single $\lcut$-node and two axiom links.
    Then $\pi'$ is a single axiom link:
    $$
    \vcenter
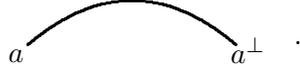
{
      \begin{xy}
        0;<2em,0ex>:<0em,4ex>::
        (1,1)="a"*{\sa}, 
        (5,1)="d"*{\sna}, 
        "a"*{\ptens};"d"*{\ptens} **\crv{~*=<.5pt>{.}(3,\doubleheight)},
      \end{xy}
      }
    \quadfs
    $$
    Identity maps are representatives of axiom links (and in
    particular of \(\Gflat(\pi')\)), and Remark~\ref{rmrk:onecutelim}
    tells us that \(\Gflat(\pi)\) has representative
    \(\id_{G(a)}\circ\id_{G(a)}\).
  \item The net $\pi'$ is obtained from $\pi$ by reducing 
    a cut formula $(A\ltens
    B)\lcut(B\lneg\lpar A\lneg)$, where both, the $\lcut$-node, as
    well as its $\ltens$-child are splitting, \ie by removing them
    \(\pi\) falls into three components:
    \begin{itemize}
    \item \sloppy First, we have the net $\pi_1$ with conclusions
      $\Gamma,A$. Let $f_1:G(\Gamma)\lneg\to G(A)$ be an arrow that
      represents $\Gflat(\pi_1)$.
    \item Second, we have the net $\pi_2$ with conclusions $\Delta,B$.
      Let $f_2:G(\Delta)\lneg\to G(B)$ represent $\Gflat(\pi_2)$.
    \item Finally, we have the net $\pi_3$ with conclusions
      $\Theta,B\lneg\lpar A\lneg$. Let $f_3:G(\Theta)\lneg\to
      G(B)\lneg\lpar G(A)\lneg$ represent $\Gflat(\pi_3)$. The same
      arrow also represents $\Gflat(\pi_4)$, where $\pi_4$ is the net
      with conclusions $\Theta,B\lneg,A\lneg$ that is obtained from
      $\pi_3$ by removing the $\lpar$.
    \end{itemize}
    Obviously the composite
      $$
      \xymatrix{ \bigl(G(\Gamma)\lneg\ltens G(\Delta)\lneg\bigr)\ltens
        G(\Theta)\lneg \ar[d]^{(f_1\ltens f_2)\ltens f_3} \\
        \bigl(G(A)\ltens G(B)\bigr)\ltens\bigl(G(B)\lneg\lpar G(A)\lneg\bigr)
        \ar[d]^{\idh\lneg_{G(A\ltens B)}} \\
        \lbot
      }
      $$
      represents $\Gflat(\pi)$.  But we can also take our three
      nets and do two tensor introductions on them, to get
      a net with conclusions \(\Delta, B\ltens B\lneg, A\lneg \ltens
      A , \Gamma, \Theta\). Let
      \[
      \xymatrix{
         G(\Delta)\lneg \ltens G(\Theta)\lneg \ltens G(\Gamma)\lneg
        \ar[r]^-{h} &
        \bigl(G(B)\ltens G(B)\lneg\bigr)  \lpar \bigl(G(A)\lneg\ltens G(A)\bigr)  
      }
      \]
      represent that net. It should be obvious that the composite
      \[
      \xymatrix{
        G(\Delta) \ltens G(\Theta)\ltens G(\Gamma) \ar[d]^-{h} \\ 
         \bigl(G(B)\ltens G(B)\lneg\bigr)  \lpar \bigl(G(A)\lneg\ltens G(A)\bigr)  
        \ar[d]^-{\idh\lneg_{G(B)}\lpar \idh\lneg_{G(A)}} \\
        \lbot \lpar \lbot \ar[d]^-{\isom}\\ \lbot
      }
      \]
      represents the net \(\Gflat(\pi')\).
      Now look at the following diagram
      {\small
      \[
      \qqquad\qqlapm{
        \xymatrix@C=4.2em@R=5ex{ G(\Gamma)\lneg \ltens G(\Delta)\lneg
        \ltens G(\Theta)\lneg \ar[r]^-{f_{1}\ltens f_{2} \ltens f_{3}}
        \ar[dd]_{\isom} & G(A)\ltens G(B) \ltens \bigl( G(B)\lneg
        \lpar G(A)\lneg \bigr) \ar[d]_{\isom}
        \ar[r]^-{\idh\lneg_{G (A\ltens B)}} & \lbot\ar[dd]^{\isom}\\
        & G(B)\ltens\bigl( G(B)\lneg \lpar G(A)\lneg\bigr) \ltens G(A) 
        \ar[d]_{w} \\
        G(\Delta)\lneg \ltens G(\Theta)\lneg \ltens G(\Gamma)\lneg
        \ar[r]_-{h} & \bigl(G(B)\ltens G(B)\lneg \bigr) \lpar \bigl(
        G(A)\lneg \ltens G(A) \bigr) 
        \ar[r]_-{\idh\lneg_{G(\!B\!)}\!\lpar\!\idh\lneg_{G(\!A\!)}} &
        \lbot \lpar \lbot }
      }
      \]
      }
      where \(w\) is
      \[
      \xymatrix{ G(B)\ltens \bigl( G(B)\lneg \lpar G(A)\lneg \bigr)
        \ltens G(A)
        \ar[d]^*+{\id_{G(B)}\ltens
          \tau_{G(B)\lneg,G(A)\lneg,G(A),\emptyset}} \\
        G(B)\ltens \bigl(G(B)\lneg \lpar( G(A)\lneg \ltens G(A))
        \bigr)
        \ar[d]^*+{\tau_{\emptyset,G(B),G(B)\lneg,G(A)\lneg\ltens G(A)}} \\
        \bigl(G(B)\ltens G(B)\lneg\bigr) \lpar \bigl( G(A)\lneg \ltens
        G(A)\bigr) }
      \]
      The left rectangle of the big diagram commutes because we can
      apply the general version of Remark~\ref{rmrk:tensorintro}
      twice, once for each occurrence of \(\tau\) in \(w\) (i.e., once
      for each tensor introduction). Then if we take the dual of \(w\), we
      get (using Equations~(\ref{eqn:switchduality}))
      \[
      \xymatrix{ \bigl( G(A)\lneg \lpar G(A)\bigr)\ltens
        \bigl(G(B)\lpar G(B)\lneg\bigr)
        \ar[d]^*+{\tau_{\emptyset,G(A)\lneg\lpar G(A),G(B),G(B)\lneg}}\\
        \bigl((G(A)\lneg \lpar G(A)) \ltens G(B)\bigr) \lpar G(B)\lneg
        \ar[d]^*+{\tau_{G(A)\lneg ,G(A),G(B),\emptyset}
          \lpar\id_{G(B)\lneg}} \\
        G(A)\lneg \lpar \bigl( G(A)\ltens G(B)\bigr) \lpar G(B)\lneg }
      \]
      and this shows that \(w\lneg =
      \tau_{G(A)\lneg,G(A),G(B),G(B)\lneg}\) because of
      Proposition~\ref{prop:twotausareone}. We can now see that the
      right half of the big diagram is exactly the dual of
      Equation~(\ref{eqn:namestensor}), thus showing that the whole
      diagram commutes, from which we get that \(\Gflat(\pi) =
      \Gflat(\pi')\).
      
    \item If none of the three cases above holds, then $\pi$ must
      contain a $\ltens$-root or a \(\lcut\) which is splitting. The
      same node is also splitting in $\pi'$. By removing it, the net
      $\pi$ falls into two parts, say $\pi_1$ and $\pi_2$. Similarly,
      $\pi'$ falls into $\pi_1'$ and $\pi_2'$.  \Wolg, we can assume
      that $\pi_1$ contains the induction hypothesis' cut. Then, we
      have that $\pi_1'$ is the result of reducing it, and also that
      $\pi_2'=\pi_2$.  We can therefore apply the induction
      hypothesis. \qed
  \end{enumerate}
We can now proceed in the proof of Theorem~\ref{thm:freestar} by showing how
the functor $G$ is defined on the arrows. Recall that each proof graph
$\prfgrf{P}{\Gamma}$ can be seen as an ordinary proof net with
conclusion $P\lfmark,\Gamma$ (see Observation~\ref{obs:unitfree}), to which we
can apply the construction of the equivariant families. This construction
gives us in particular for each proof graph $\prfgrf{P}{A\lneg,B}$ a unique
arrow $\psi_{\prfgrf{P}{A\lneg,B}}:G(P\lfmark)\lneg\to G(A\lneg)\lpar G(B).$

Furthermore, every linking $P$ uniquely determines an arrow
$\phi_P:\loneC\to\Gbn{P}$ in $\CC$, which is inductively obtained as follows:
$$
\begin{array}{l@{\;=\;}l@{:\lone\to\;}l}
  \phi_\lone&\id_\lone&\lone\\
  \phi_{a\ltens a\lneg}&\idh_{G(a)\lneg}&G(a)\lpar G(a)\lneg\\
  \phi_{a\lneg\ltens a}&\idh_{G(a)}&G(a)\lneg\lpar G(a)\\
  \phi_{\lbot\ltens P'}&\rho\lneg_{\Gbn{{P'}}}\fcomp\phi_{P'}&
  \lbot\lpar\Gbn{{P'}}\\
  \phi_{P'\ltens\lbot}&\lambda\lneg_{\Gbn{{P'}}}\fcomp\phi_{P'}&
  \Gbn{{P'}}\lpar\lbot\\
  \phi_{P'\lpar P''}&(\phi_{P'}\ltens\phi_{P''})\fcomp\lambda_\lone^{-1}&
  \Gbn{{P'}}\ltens\Gbn{{P''}}
\end{array}
$$ The arrow $\phi_P$ can be composed with $\psi_{\prfgrf{P}{A\lneg,B}}$ to
get $\xi_{\prfnet{P}{A\lneg,B}}:\lone\to G(A\lneg)\lpar G(B)$.  That this is
well-defined, is ensured by the following lemma (in which we finally use 
the fact that the units are units).

\begin{lemma}\label{lem:unique}
If $\prfgrf{P}{A\lneg,B}\prfequ\prfgrf{Q}{A\lneg,B}$, then 
$\xi_{\prfnet{P}{A\lneg,B}}=\xi_{\prfnet{Q}{A\lneg,B}}$.
\end{lemma}

\proof
  First notice that all the one-step equivalences in~\ref{par:grfequ}
  involve two formulas that have exactly the same set of atoms (here,
  naturally, \(\lone\) and \(\lbot\) are considered to be atoms). 
  \Wolg, let us assume that  \(P\) is the linking tree of
  the left-hand side of one of these equivalences
  and \(Q\) is the linking tree of the right-hand side. 
  Then there is an ordinary proof net for
  the sequent \(P\lfmark,Q\lfmarklneg\), in which the graph of axiom
  links forms a bijection between the atoms of the two formulas.\footnote{%
    In all cases except the last one, it does not matter whether $P\lfmark$ or
    $Q\lfmark$ is negated, but in the last one (the one with the side
    condition) we only get a correct ordinary proof net if we negate
    $Q\lfmark$.}
  Let us call this
  proof net \(\pi_1\). Thus \(\Gflat(\pi_1)\) defines a map
  \(\theta\colon G(Q\lfmark) \to G(P\lfmark)\). In addition, \(\theta\) is
  always an isomorphism in \(\CC\). This is because everything that has to do
  with the units ``just melts away'' and the two formulas
  $\Gbn{Q}\isom\Gbn{P}$ are thus both isomorphic to $\lbigtens{a\ltens
    a\lneg}{\set{G(a)\lpar G(a)\lneg}}$, where $a\ltens a\lneg$ ranges
  over the ``real'' axiom links, which are the same in $Q$ and $P$.
  Therefore we immediately have that the left-hand side triangle in
  the following diagram commutes:
  $$
  {
    \phantom{\strut\lone}
    
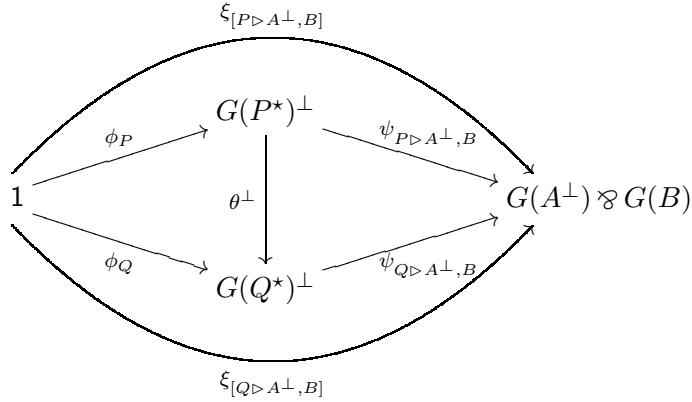
\xymatrix@C=6em@R=3ex{
      &\Gbn{P}\ar[dr]^{\psi_{\prfgrf{P}{A\lneg,B}}}\ar[dd]_{\theta\lneg}\\
      {\qquad\strut\lone}\ar[ur]^{\phi_P}\ar[dr]_{\phi_Q}
      \ar@/^13ex/[rr]^{\xi_{\prfnet{P}{A\lneg,B}}}
      \ar@/_13ex/[rr]_{\xi_{\prfnet{Q}{A\lneg,B}}}
      &&
      \rlapm{G(A\lneg)\lpar G(B)}\phantom{\strut\lone\quad\qquad}\\
      &\Gbn{Q}\ar[ur]_{\psi_{\prfgrf{Q}{A\lneg,B}}}\\
    }\phantom{G(A\lneg)\lpar G(B)}
  }
  $$
  For showing that the right-hand side triangle commutes we will apply
  Lemma~\ref{lem:unitfree}. For this, consider the following four ordinary
  proof nets: 
  \begin{enumerate}
  \item \(\pi_1\) is as above.
  \item $\pi_2=\unitfree{P}{A\lneg,B}$ 
    is the ordinary proof net with conclusions
    $P\lfmark,A\lneg,B$ that corresponds to the proof graph
    $\prfgrf{P}{A\lneg,B}$ (see Observation~\ref{obs:unitfree}).
  \item Similarly, $\pi_3=\unitfree{Q}{A\lneg,B}$ is the ordinary proof
    net with conclusions $Q\lfmark,A\lneg,B$ that corresponds to the proof
    graph $\prfgrf{Q}{A\lneg,B}$.
  \item Finally, the ordinary proof net $\pi_4$ is obtained from
    $\pi_1$ and $\pi_3$ by connecting $Q\lfmarklneg$ and $Q\lfmark$
    with a $\lcut$-node.
  \end{enumerate}
  By definition, the arrows $\psi_{\prfgrf{P}{A\lneg,B}}$ and
  $\psi_{\prfgrf{Q}{A\lneg,B}}$ are representatives of the equivariant
  families obtained from $\pi_2$ and $\pi_3$, respectively. Similarly,
  the isomorphism $\theta\lneg$ is a representative of the equivariant
  family obtained from $\pi_1$. Consequently, the composition
  $\psi_{\prfgrf{Q}{A\lneg,B}}\fcomp\theta\lneg$ is a representative
  of the equivariant family obtained from $\pi_4$ (because of
  Remark~\ref{rmrk:onecutelim}).  Furthermore, it is easy to see that
  eliminating the cut from $\pi_4$ (as defined in
  \ref{par:unitfreecutelim}) yields $\pi_2$. Therefore we can apply
  Lemma~\ref{lem:unitfree} to get that
  $\psi_{\prfgrf{Q}{A\lneg,B}}\fcomp
  \theta\lneg=\psi_{\prfgrf{P}{A\lneg,B}}$.  \qed

We have shown that for any proof net of the form
$f=\prfnet{P}{A\lneg,B}$, the arrow $G(f):G(A)\to G(B)$ determined by
$\xi_{\prfnet{P}{A\lneg,B}}$ via Proposition~\ref{prop:multi} is
uniquely defined.  It remains to prove that $G:\PN(\cA)\to\CC$ is
indeed a functor (\ie identities and composition are preserved).  That
for each formula $A$, the proof $\prfnet{I_A}{A\lneg,A}$ is mapped to
identity $\id:G(A)\to G(A)$ is an easy induction on the structure of
$A$ and left to the reader.  The crucial part is to show that for two
given proof nets $f=\prfnet{P}{A\lneg,B}$ and
$g=\prfnet{Q}{B\lneg,C}$, the composition $G(g)\fcomp G(f)$ yields the
same arrow in $\CC$, as $G(g\fcomp f)$.

Observe that $g\fcomp f=\prfnet{R}{A\lneg,C}$ is the proof net that
is obtained by eliminating the cut in $\prfnet{P\lpar Q}{A\lneg,B\lcut
B\lneg,C}$, whose equivariant family is determined by
\[
\vcenter{\xymatrix{
  \lone \ar[d]^{\isom} \\
  \lone\ltens\lone \ar[d]^{\phi_{P} \ltens \phi_{Q}} \\
  \Gbn{P} \ltens \Gbn{Q} \ar[d]^{\psi_{\prfgrf{P}{A\lneg \lpar B}} \ltens
    \psi_{\prfgrf{Q}{B\lneg \lpar C}}} \\
  \bigl(G(A)\lneg \lpar G(B)\bigr)\ltens\bigl(G(B)\lneg\lpar G(C)\bigr) 
  \ar[d]^{\tau_{G(A)\lneg,G(B),G(B)\lneg,G(C)}} \\
  G(A)\lneg \lpar\bigl(G(B)\ltens G(B)\lneg\bigr)\lpar G(C) 
  \ar[d]^{\id \ltens \idh\lneg \ltens \id}\\
  G(A)\lneg \lpar G(C) }}
\quadcm
\]
and a by-now standard argument tells us that this is \(\widehat{G(g)
  \circ G(f)}\).  In order to show that this is also
\(\widehat{G(g\circ f)}\), it suffices to show the following general
result.

\begin{lemma}\label{lem:comppres}
  Let $\prfgrf{T}{\Gamma}\prfred\prfgrf{S}{\Delta}$, \ie the proof
  graph $\prfgrf{S}{\Delta}$ is obtained from $\prfgrf{T}{\Gamma}$ by
  applying a single cut reduction step. Then
  $\xi_{\prfnet{T}{\Gamma}}$ and $\xi_{\prfnet{S}{\Delta}}$ denote the
  same morphism $\lone_\CC\to\lbigpar{}{\set{G(A_1),\ldots,G(A_n)}}$,
  where $A_1,\ldots,A_n$ are the formulas in $\Gamma$ (resp.~$\Delta$)
  that are not cuts.
\end{lemma}

\proof
  The proof is very much like that of Lemma~\ref{lem:unique}: for
  every case of a cut reduction step we will construct an ordinary
  proof net \(\pi_{1}\) whose conclusions will involve \(T\lfmark,
  S\lfmarklneg\), and which will also define a map \(\theta \colon
  G(S\lfmark)\to G(T\lfmark)\). Then we will show that the two
  triangles below commute:
$$
{
\phantom{\strut\lone}
\xymatrix@C=6em@R=2ex{
&\Gbn{T}\ar[dr]^{\psi_{\prfgrf{T}{\Gamma}}}\ar[dd]_{\theta\lneg}\\
{\qquad\strut\lone}
\ar[ur]^{\phi_T}\ar[dr]_{\phi_S}
&&
\rlapm{\lbigpar{}{\dset{G(A_1),\ldots,G(A_n)}}}\phantom{\strut\lone\qquad}\\
&\Gbn{S}\ar[ur]_{\psi_{\prfgrf{S}{\Delta}}}\\
}\phantom{\lbigpar{}{\dset{G(A_1),\ldots,G(A_n)}}}
}
$$
Again we need three ordinary proof nets in addition of \(\pi_{1}\):
\begin{enumerate}
\stepcounter{enumi}
\item Let $\pi_2=\unitfree{T}{\Gamma}$ with conclusions
  $T\lfmark,A_1,\ldots,A_n,B_1\lcut B_1\lneg,\ldots,B_m\lcut
  B_m\lneg$, where $B_1\lcut B_1\lneg,\ldots,B_m\lcut B_m\lneg$ are
  the cuts in $\Gamma$. Applying the construction of the equivariant
  family yields the arrow $\psi_{\prfgrf{T}{\Gamma}}$.
\item Let $\pi_3=\unitfree{S}{\Delta}$ with conclusions
  $S\lfmark,A_1,\ldots,A_n,C_1\lcut C_1\lneg,\ldots,C_l\lcut
  C_l\lneg$, where $C_1\lcut C_1\lneg,\ldots,C_l\lcut C_l\lneg$ are
  the cuts in $\Delta$. The arrow $\psi_{\prfgrf{S}{\Delta}}$ is
  obtained by applying the construction of the equivariant family to
  $\pi_3$.
\item Finally, $\pi_4$ is obtained by composing $\pi_1$ and $\pi_3$
  with a cut on $S\lfmarklneg$ and $S\lfmark$.
\end{enumerate}
There are three cases to consider:
\begin{itemize}
\item The reduced cut (see Section~\ref{sec:cutelim}) is on binary
  connectives. In this case $T\lfmark$ and $S\lfmark$ are identical,
  and $\pi_1$ is the usual identity net, so that \(\theta\) is the
  identity map.  The commutativity of the left-hand side triangle
  follows trivially. For the right-hand side triangle, observe that
  the result of eliminating the \(S\lfmark\)-cut on the composite
  \(\pi_{4}\) yields exactly \(\pi_{2}\). We can therefore apply
  Lemma~\ref{lem:unitfree}.
\item The reduced cut is on atoms. Then (see
  Section~\ref{sec:cutelim}) we can assume that
  $T\lfmark=P\cons{(a_k\lneg\ltens a_j)\lpar(a_i\lneg\ltens a_h)}$ and
  $S\lfmark=P\cons{a_k\lneg\ltens a_h}$ for some context $P\conhole$.
  Furthermore, one of the cuts in $\Gamma$ is $a_i\lcut a_j\lneg$. It
  should be clear that there is a correct ordinary proof net \(\pi_0\)
  with conclusions
  \[
  (a_k\lneg\ltens a_j)\lpar(a_i\lneg\ltens a_h),\:
  a_h\lneg \lpar a_k,\: a_i\lcut a_j\lneg
  \quadcm
  \]
  where the same index on an atom and a negated atom denotes the
  presence of an axiom link between the two (here there is a little
  breach in our previous convention of using a different index for
  every single atom). Out of this we can construct \(\pi_{1}\) with
  conclusions
  \[
  P\cons{(a_k\lneg\ltens a_j)\lpar(a_i\lneg\ltens
    a_h)},\:P\lneg\cons{a_h\lneg\lpar a_k},\:a_i\lcut a_j\lneg
  \quadcm
  \]
  where the additional axiom links simply connect every atom of the
  context \(P{\conhole} \) to its corresponding negation in
  \(P\lneg{\conhole}\). It should be clear that \(\pi_{1}\) is
  correct.  Furthermore, the result of eliminating the two cuts
  $S\lfmarklneg\lcut S\lfmark$ and $a_i\lcut a_j\lneg$ in the
  composite \(\pi_{4}\) yields exactly \(\pi_{2}\). Therefore, by
  Lemma~\ref{lem:unitfree}, the right-hand side triangle commutes.
  For the commutativity of the left-hand side triangle consider again
  the net $\pi_0$. A representative for the equivariant family
  \(\Gflat(\pi_0)\) is given by the map
  \[
  \xymatrix{ 
    \bigl(G(a_{h})\lneg \lpar G(a_{i})\bigr) \ltens
    \bigl(G(a_{j})\lneg \lpar G(a_{k})\bigr) 
    \ar[d]^-{\tau} \\
    G(a_{h})\lneg\lpar\bigl(G(a_{i})\ltens G(a_{j})\lneg\bigr) 
    \lpar G(a_{k})
    \ar[d]^-{\id\lpar\idh\lneg\lpar\id} \\
    G(a_{h})\lneg \lpar G(a_{k}) }
    \]
    (here the indices are used only as position markers; there are
    only two distinct ``atomic'' objects of \(\CC\), namely $G(a)$ and
    $G(a)\lneg$). By definition, this is the
    internal composition
    \[
    \xymatrix{ \bigl(G(a)\lneg \lpar G(a)\bigr)\ltens \bigl(G(a)\lneg
      \lpar G(a)\bigr) \ar[rrr]^-{\gamma_{G(a),G(a),G(a)}} &&& 
      G(a) \lneg \lpar G(a)\quadfs  }
    \]
    If we compose this with the names of the identity: 
    {\small
      $$
      \qqquad\qqlapm{
        \xymatrix{ \lone \ar[r]^-{\isom} &\lone\ltens\lone
        \ar[rr]^-{\idh_{G(a)}\ltens \idh_{G(a)}} && \bigl(G(a)\lneg
        \lpar G(a)\bigr)\ltens \bigl(G(a)\lneg \lpar G(a)\bigr)
        \ar[rr]^-{\gamma_{G(a),G(a),G(a)}} && G(a) \lneg \lpar G(a) }}
      $$
      }%
    we get (by Proposition~\ref{prop:namescompose}) the name of the
    identity \(\idh_{G(a)}\colon \lone \to G(a) \lneg \lpar G(a) \),
    which in turn represents the result of eliminating the cut from
    \(\pi_0\).  Therefore the left-hand side triangle commutes for
    the case where $P\conhole=\conhole$ is the empty context. The
    general case follows by a straightforward induction on
    $P\conhole$.
  \item The reduced cut is on units. Then we have that
    $T\lfmark=P\cons{\lbot_j\lpar(\lone_i\ltens Q)}$ and
    $S\lfmark=P\cons{Q}$ for some context $P\conhole$ and some formula
    $Q$ (and one of the cuts in $\Gamma$ is $\lbot_i\lcut\lone_j$). We
    now construct $\pi_1$ so that its conclusions are
    \[
    P\cons{\lbot_j\lpar(\lone_i\ltens Q)},\:P\lneg\cons{Q\lneg},\:
    \lbot_i\lcut\lone_j \quadcm
    \]
    in other words, in such a way that the only non-trivial axiom
    links are indicated by the indices $i$ and $j$. It should be
    obvious that this gives a correct (ordinary) net. We now see that
    the map \(\theta\) is an isomorphism and that the two triangles
    commute, since all the syntactical entities that do not belong to
    $P\conhole$ and $Q$ simply ``melt away'' in the categorical
    interpretation because they follow the coherence laws for
    units.\qed
  \end{itemize}

We now have completed the proof of Theorem~\ref{thm:freestar}.
To see this, let us recall the main features of our construction:
\begin{enumerate}
\item Functoriality of $G$ together with Lemma~\ref{lem:unique} and
  Proposition~\ref{prop:multi} ensure that $G$ does preserve the
  *-autonomous structure.
\item By sequentialisation, every morphism in the category $\PN(\cA)$ is
  expressible in the terms of the *-autonomous structure imposed on $\PN(\cA)$
  in Section~\ref{sec:PNstar}. Consequently, the functor $G$ is indeed
  uniquely defined by the data given in Theorem~\ref{thm:freestar}.
\end{enumerate}

\bigskip

It might be worth mentioning, that the main result of
\cite{balat:dicosmo:99}, namely that two MLL formulas are isomorphic
if and only if they can be transformed into each other by applying the
standard rewriting rules of associativity, commutativity, and unit
(for $\lpar$ and $\ltens$), is an immediate consequence of
Theorem~\ref{thm:freestar}. Furthermore,  Theorem~\ref{thm:freestar}
provides a decision procedure for the equality of morphisms in the
free symmetric *-autonomous category, which is in our opinion
simpler than the ones provided in \cite{blute:etal:96} and
\cite{koh:ong:99}.

\section{Conclusion}

We think we made a convincing case for the the cleanest approach yet
to proof nets with the multiplicative units. In particular, our main
results are stated in such a way as to be easily applicable; in
addition our techniques can certainly be used in more general
situations than purely multiplicative linear logic.

We began with a discussion on the relationship between proof systems
and categories; it turns out that the writing up of this paper gave
many occasions to deepen that reflection, and more will said about
that relationship in subsequent work. We made use of two unstated
assumptions, which certainly belong to ``mainstream ideology'':
\begin{itemize}
\item that there is a single way to introduce bottom (for instance we
  also could have a special axiom for it),
\item that the standard equations for units in monoidal categories
  should be used for proofs.
\end{itemize}
We now think that these standard postulates deserve more
scrutiny~\cite{lam:str:05:naming,lam:str:05:freebool}, but we make no
predictions about the conclusions we will eventually reach.

These new subtleties in no way modify our general belief: that
category theory should be used as a general algebraic yardstick for
tackling the questions related to identifications of proofs. As we
have said at the beginning, this can work well only if we allow a
certain ideological flexibility on \emph{both} the proof-theoretical
and category-theoretical side.

There are some issues that are left open and that we want to explore
in the future:
\begin{itemize}
\item The addition of Mix. Ordinary proof nets have a weaker version
  of the Danos-Regnier correctness criterion (every switching produces
  an acyclic, but not necessarily connected, graph) which gives a
  sequentialization theorem for MLL with the (binary) Mix rule
  added~\cite{fleury:retore:94}. An important property of this setting
  is that when a net is correct, the number of connected components is
  invariant with respect to the actual switching.  It is not hard to
  see that if we add binary Mix to our theory of proof nets, this
  invariant is respected by both bottom introductions and the
  equations of~\ref{par:grfequ}. This allows us to say that our theory
  extends to MLL with Mix, although there is some nontrivial work left
  to be done, namely to prove that we actually have constructed the
  free *-autonomous category with Mix. Thus we have to manage the
  additional algebra needed for this
  (cf.~\cite{cockett:seely:97:mix,fuhrmann:pym:dj,lam:str:05:freebool}),
  which involves the necessary equations that are required to obtain a
  coherence result when the ``mother of all mix maps'' \(\lbot \to
  \lone\) is added to a *-autonomous category. Another, also standard
  view of Mix is adding the requirement that this map be an
  isomorphism. But then the theory of proof nets presented
  in~\cite{fleury:retore:94} (when we add one constant with two
  introduction rules to the logic) is sufficient to deal with this
  case.
\item Exploring the noncommutative world, especially the particular
  logic where the context structure is no longer a multiset of
  formulas but a cyclic order of formulas
  \cite{yetter:90,lamarche:retore:96}.  In the unit-free case, the
  correctness criterion has to be modified such that the net has to be
  planar (i.e., no crossings of edges are allowed).  It is easy to see
  that our correctness criterion and the equivalence relation defined
  in \ref{par:grfequ} can be adapted accordingly. However, the
  question is whether we can obtain a well-behaved cut elimination
  such that we can construct the free cyclic *-autonomous
  category \cite{rosenthal:Bimod,barr:95,schneck:99}. Here is another
  interesting question: Could it be that in the noncommutative case we
  can find normal representatives for proofs instead of having to rely
  on equivalence classes?
\item The relation with the calculus of
  structures~\cite{guglielmi:strassburger:01,brunnler:tiu:01} and its
  use of deep inference.  We should mention that the idea behind our
  approach originates from the new viewpoints that are given by deep
  inference.
\item The addition of additives to our theory. This should not be
  very hard, given the work done in~\cite{hughes:glabbeek:03}. The
  true challenge is to include also the additive units.
\item The development of a theory of proof nets for classical logic.  The
  problem is finding the right extension of the axioms of a *-autonomous
  category, such that on one hand classical proofs are identified in a natural
  way, but that on the other hand there is no collapse to a boolean
  algebra. While we were writing this, we became aware
  of~\cite{fuhrmann:pym:04,fuhrmann:pym:oecm,fuhrmann:pym:dj}, which tackle
  this very problem.  Some additional
  research~\cite{lam:str:05:naming,lam:str:05:freebool,lamarche:gap,str:medial}
  allows us to say that the last word on the relationship between classical
  logic and categories will not be said in the near future.
\item The search for meaningful invariants. It is very probable that
  the equivalence classes of graphs we define have a geometric
  meaning, and can be related to more abstract invariants like those
  given by homological algebra. We are convinced that the work in
  in~\cite{metayer:94} is only the tip of the iceberg.
\end{itemize}

\bibliographystyle{alpha}
\bibliography{my_lit}

\end{document}